\documentclass[journal]{IEEEtran}
\usepackage{amssymb}
\usepackage{amsmath,amsthm,amsfonts,amssymb,bm}
\usepackage{algorithmic}
\usepackage{algorithm}
\usepackage{array}
\usepackage{textcomp}
\usepackage{breakurl}
\usepackage{pifont}
\usepackage{stfloats}
\usepackage{url}
\usepackage{verbatim}
\usepackage{graphicx}
\usepackage{subfigure}
\usepackage{cite}
\usepackage{makecell}
\usepackage{multirow}
\usepackage{booktabs}
\usepackage{tablefootnote}
\usepackage{color}

\begin{document}
\title{{MetaLoc: Learning to Learn Wireless Localization}
\author{Jun Gao,~\IEEEmembership{Graduate Student Member,~IEEE}, Dongze Wu, Feng Yin,~\IEEEmembership{Senior Member,~IEEE}, Qinglei Kong,~\IEEEmembership{Member,~IEEE}, Lexi Xu,~\IEEEmembership{Senior Member,~IEEE}, Shuguang Cui,~\IEEEmembership{Fellow, IEEE}}
\thanks{This paper is an extension of our work~\cite{gao2022metaloc} presented in the proceedings of the IEEE International Conference on Communications (ICC), Seoul, Korea, in May 2022. Corresponding author is Feng Yin (yinfeng@cuhk.edu.cn).}
\thanks{J. Gao is with the Future Network of Intelligence Institute (FNii), and also with the School of Science and Engineering (SSE), The Chinese University of Hong Kong, Shenzhen, China (e-mail: jungao@link.cuhk.edu.cn).}
\thanks{D. Wu was with the School of Science and Engineering (SSE), The Chinese University of Hong Kong, Shenzhen, China, and is now with the Department of Statistics, University of Oxford, United Kingdom  (e-mail: dongzewu@link.cuhk.edu.cn).}
\thanks{F. Yin is with the School of Science and Engineering (SSE), The Chinese University of Hong Kong, Shenzhen, China (e-mail: yinfeng@cuhk.edu.cn).}
\thanks{Q. Kong is with the Institute of Space Science and Applied Technology, Harbin Institute of Technology (Shenzhen), and also with the Guangdong Provincial Key Laboratory of Future Networks of Intelligence, Shenzhen, China (e-mail: kql8904@163.com).}
\thanks{L. Xu is with the Research Institute, China United Network Communications Corporation, Beijing, China (e-mail: davidlexi@hotmail.com\& xulx29@chinaunicom.cn).}
\thanks{S. Cui is with the School of Science and Engineering (SSE), the Future Network of Intelligence Institute (FNii), and the Guangdong Provincial Key Laboratory of Future Networks of Intelligence, The Chinese University of Hong Kong; he is also affiliated with Peng Cheng Laboratory, Shenzhen, China (e-mail: shuguangcui@cuhk.edu.cn).}}

\maketitle
\pagestyle{plain}
\thispagestyle{plain}
\begin{abstract}
Existing localization methods that intensively leverage the environment-specific received signal strength (RSS) or channel state information (CSI) of wireless signals are rather accurate in certain environments. However, these methods, whether based on pure statistical signal processing or data-driven approaches, often struggle to generalize to new environments, which results in  considerable time and effort being wasted. To address this challenge, we propose MetaLoc, which is the first fingerprinting-based localization framework that leverages the Model-Agnostic Meta-Learning (MAML). Specifically, built on a deep neural network with strong representation capabilities, MetaLoc is trained on historical data sourced from well-calibrated environments, employing a two-loop optimization mechanism to obtain the meta-parameters. These meta-parameters act as the initialization for quick adaptation in new environments, reducing the need for much human effort. The framework introduces two paradigms for the optimization of meta-parameters: a centralized paradigm that simplifies the process by sharing data from all historical environments, and a distributed paradigm that maintains data privacy by training meta-parameters for each specific environment separately. Furthermore, the advanced distributed paradigm modifies the vanilla MAML loss function to ensure that the reduction of loss occurs in a consistent direction across various training domains, thus facilitating faster convergence during training. Our experiments on both synthetic and real datasets demonstrate that MetaLoc outperforms baseline methods in terms of localization accuracy, robustness, and cost-effectiveness. The code and datasets used in this study are publicly available\footnote{Codes and datasets can be found at:\url{https://github.com/WU-Dongze/MetaLoc}}. 
\end{abstract}

\begin{IEEEkeywords}CSI, meta-learning, RSS, sample efficiency, wireless localization.
\end{IEEEkeywords}

\section{Introduction}

Location-based services have become an integral part of our daily lives, and various localization techniques have been extensively researched for over a century by various scientific communities~\cite{faragher2015location, 9992123, 6475197,godrich2010target}. Although existing global navigation satellite systems (GNSS) provide high outdoor localization accuracy, the demands of emerging applications in diverse areas, such as autonomous driving~\cite{bresson2017simultaneous}, cooperative 3D scene reconstruction~\cite{wei2022stereo}, and epidemic tracking~\cite{jeong2019smartphone}, require higher levels of accuracy, cost-effectiveness, and robustness. Moreover, with the emergence of mmWave and massive MIMO technologies in 5G and 6G communications, there is a growing demand for precise location information~\cite{zhou2019successive,gante2020deep, wang2022location}. In light of these growing needs, it is crucial to develop a state-of-the-art localization system that can cover both outdoor and complex indoor environments to achieve a full range of high-precision location-based services~\cite{zafari2019survey,ali20206g,xing2022integrated}.

The field of fingerprinting-based localization has recently gained much attention for indoor environments, and the process involves two stages: the offline stage and the online stage. During the offline stage, features of received signals at reference points (RPs) with known locations are collected to form a fingerprint database. These signals are transmitted by access points (APs) in the environment, and the representative features typically include received signal strength (RSS), channel state information (CSI), and magnetic field information. In the online stage, signal features collected at an unknown location, also known as a test point (TP), are compared with the established database to determine the predicted location using algorithms such as RADAR~\cite{bahl2000radar} and Horus~\cite{youssef2005horus}.

Wireless signal propagation is susceptible to even slight changes in the environment, such as opening a door or the presence of people moving around, which can result in inconsistent fingerprints even if a user remains in the same location. This makes it challenging to construct an accurate statistical fingerprint database that accurately represents the entire area of interest. To tackle this challenge, data-driven localization mechanisms using machine learning techniques have gained increasing attention in recent years~\cite{hsieh2019deep,yin2017distributed,jin2020bayesian}. Machine learning techniques, being data-hungry, demand large amounts of data samples. Thus, building a database for each target indoor localization becomes time-consuming and labor-intensive. To overcome this challenge, various machine learning techniques have been established and applied to indoor localization, such as data augmentation~\cite{sinha2019data}, semi-supervised learning techniques~\cite{shrivastava2017learning}, and informed machine learning~\cite{von2021informed}.

However, despite the advances made in indoor localization using machine learning, it is still a challenge to build a model that is universally applicable to all indoor environments. Most studies in the field to date have only focused on one specific environment, such as a single room or floor of a building~\cite{bahl2000radar,7827145,torres2014ujiindoorloc}. This narrow focus means that there is no guarantee that a pre-selected machine learning model that performs well in one environment will be effective in others. To deploy a machine learning model in a new environment, one often needs to collect extensive data through site surveys, rebuild the fingerprint database, and then retrain the model, which can be time-consuming and resource-intensive. This highlights the need for a machine learning model that can learn the essential channel features and be broadly applicable to all indoor environments, as pointed out in a recent 6G white paper~\cite{ali20206g}.
\subsection{Related Works}
\begin{table*}
\caption{Comparisons of different localization methods }
\centering
\begin{tabular}{cccccc} 
\toprule[1.5pt] 

\multicolumn{1}{c}{\textbf{Methods}} & \multicolumn{1}{c}{\textbf{Signal Features}} & \multicolumn{1}{c}{\textbf{Models}}& \multicolumn{1}{c}{\textbf{Accuracy}} & \multicolumn{1}{c}{\textbf{Robustness}}& \multicolumn{1}{c}{\textbf{Cost-effectiveness}}\\  
\hline 
TransLoc~\cite{Transloc} & RSS & Machine learning & 1.82 m$\sim$2.81 m & \checkmark & \ding{53} \\
ViVi \cite{wu2017gain} & RSS & Deterministic & 3.30 m$\sim$4.30 m  & \checkmark & \ding{53} \\
AcMu \cite{wu2015static} & RSS & Deterministic & 1.40 m$\sim$3.00 m  & \checkmark & \ding{53} \\
DFPS \cite{fang2015novel}& RSS & Machine Learning & 1.2m$\sim$2.8m  & \checkmark & \ding{53} \\
FILA \cite{wu2012fila}& CSI  & Deterministic & 0.45 m $\sim$ 1.2 m & \ding{53} & \ding{53} \\
DeepFi~\cite{wang2015deepfi} & CSI & Probabilistic & 0.95 m$\sim$1.80 m & \ding{53} & \checkmark\\
CiFi~\cite{cifi} & CSI & Machine learning & 1.50 m$\sim$3.00 m & \ding{53} & \checkmark\\
ConFi~\cite{chen2017confi} & CSI & Machine learning  & 1.36 m & \ding{53} & \ding{53}\\
CRISLoc~\cite{CRISLoc} & CSI & Machine learning & 0.29 m & \checkmark & \ding{53}\\
Fidora \cite{chen2022fidora} & CSI & Machine Learning & submeter-level  & \checkmark & \ding{53}\\
DAFI \cite{li2021dafi} & CSI & Machine Learning  & 97.6\% $\sim$ 89.3\% & \checkmark & \ding{53}\\
ILCL \cite{zhu2022intelligent} & CSI & Probabilistic & 1.28m$\sim$2.38m  & \checkmark & \checkmark\\

\bottomrule[1.5pt] 
\end{tabular}
\label{tab:comparison}
\end{table*}
Indoor localization techniques have been in development for many years, and the existing methods can be generally summarized from the perspective of classic signal processing (including both probabilistic and deterministic methods) and machine learning-based localization. 
\subsubsection{Probabilistic Localization}
Probabilistic localization is a method that utilizes statistical information to determine the location of a target. This is achieved by comparing the received signal measurements with a pre-built fingerprint database. One of the well-known probabilistic localization methods is Horus~\cite{youssef2005horus}, which employs a probabilistic model to characterize the signal distribution and calculates the maximum posterior probability of the target's location. Another approach, as discussed in \cite{jin2020bayesian}, involves the use of Bayesian networks for cooperative localization based on RSS. DeepFi~\cite{wang2015deepfi} further optimizes computational efficiency by combining a probabilistic model with a greedy learning algorithm. Despite having lower computational requirements, probabilistic localization methods can be challenging to implement in dynamic environments as they rely on accurate position-related measurements.

\subsubsection{Deterministic Localization}
The deterministic methods for indoor localization mainly rely on the similarity metric in the signal space to estimate the physical location of a target. This is done by determining the closest fingerprint location in the signal space, which serves as the estimated location of the target. The most widely used deterministic method is the $K$-nearest neighbors (KNN) algorithm, which considers various similarity measures between the target's signal and the fingerprints in the database. Some commonly used similarity measures include the Euclidean distance \cite{wu2017gain}, its temporal weighted version \cite{7565565}, the cosine similarity \cite{he2014sectjunction}, the Tanimoto similarity \cite{jiang2012ariel}, and others. While deterministic localization methods are relatively straightforward to implement, they can be affected by statistical fluctuations in wireless signals, which can result in a dispersed set of neighbors that are far apart in physical space, leading to less accurate localization.

 \subsubsection{Machine learning-based Localization} 
In recent years, machine learning has played a significant role in the field of localization services. Ghzali \textit{et al.} attempted to address the difficulties associated with indoor localization by approaching it as a regression problem based on RSS gathered in a real office space. Their proposed solution relied on a neural network with random initialization, which unfortunately necessitated the collection of a substantial amount of data in order to train the model effectively~\cite{ghozali2019indoor}. ConFi~\cite{chen2017confi} was the first work to explore the use of convolutional neural networks (CNNs) for learning CSI images at RP, opening up new possibilities for indoor localization. Hsieh \textit{et al.} attempted to solve the indoor localization challenge by utilizing the RSS and CSI data as a classification problem. They evaluated various neural network architectures in an effort to find the best fit for accurately estimating the location of an object within a specific room~\cite{hsieh2019deep}. Despite the potential of these machine learning-based methods, the dynamic nature of indoor environments continues to pose challenges to their robustness.

Improved robustness in localization is being achieved through both the model and data sides. On the model side, domain adaptation techniques like transfer learning are widely used, where a source domain is the original environment and the target domain is a new and potentially unseen environment. For example, TransLoc~\cite{Transloc} uses transfer learning to find the appropriate cross-domain mappings and create a homogeneous feature space that contains discriminative information from different domains. CRISLoc~\cite{CRISLoc} also uses transfer learning to reconstruct a high-dimensional CSI fingerprint database based on outdated fingerprints and a few new measurements. Fidora~\cite{chen2022fidora} trains a domain-adaptive classifier that adjusts itself to new data using a variational autoencoder, as well as a joint classification and reconstruction structure. ILCL~\cite{zhu2022intelligent} uses incremental learning and expands neural nodes for adaptation and reduced training time, but can still overfit with small numbers of CSI images. On the data side, ViVi~\cite{wu2017gain} reduces uncertainty in RSS fingerprints by exploiting spatial gradients among multiple locations. CiFi~\cite{cifi} uses phase differences between antenna pairs instead of raw measurements to improve the stability of CSI fingerprints. DFPS~\cite{fang2015novel} combines raw RSS and the difference between AP pairs to enhance robustness against heterogeneous hardware.

Despite the numerous efforts to identify cost-effective alternatives to traditional site surveys for indoor localization, challenges persist. The semi-supervised learning technique~\cite{shrivastava2017learning}, which incorporates a limited number of labeled data samples with an abundant amount of low-cost unlabeled data, has been proposed as a potential solution. However, it still struggles to mitigate the challenge of learning the noise information in the data rather than the true patterns. The informed machine learning approach~\cite{von2021informed,deist2019simulation}, which integrates data and prior knowledge, is another effort to enrich the information contained in the training data, but the use of computer simulation results as a knowledge representation can only do so much. Meanwhile, crowdsourcing and federated learning-based
approaches~\cite{yang2013freeloc,yin2020fedloc} allow for decomposing large-scale fingerprint
collections into smaller and local tasks, enabling mobile
users to participate using heterogeneous devices, but they struggle to deal with the inconsistencies of fingerprints over space and time. The difficulties in overcoming these challenges highlight the ongoing need to find a cost-effective indoor fingerprinting-based localization method with fewer site surveys.

Table~\ref{tab:comparison} provides a  comparison of various indoor localization methods in terms of accuracy, robustness, and cost-effectiveness. It is worth noting that the reported localization accuracy in different studies cannot be fairly compared due to the different datasets utilized.

\begin{table*}
\caption{Important notations used throughout the paper}
\centering
\begin{tabular}{cc} 
\toprule[1.5pt] 
\multicolumn{1}{c}{\textbf{Notation}}& \multicolumn{1}{c}{\textbf{Description}}\\  
\hline 
\multicolumn{1}{c}{$\mathcal{P}(\tau)$}                     & \multicolumn{1}{c}{The overall distribution of tasks}          \\
\multicolumn{1}{c}{$\mathcal{P}^{\left(i\right)}(\tau)$}                     & \multicolumn{1}{c}{Distribution of tasks from domain $i$}           \\
\multicolumn{1}{c}{$f_{\bm{\theta}}$}                     & \multicolumn{1}{c}{Neural Network with parameters $\bm{\theta}$}  \\
\multicolumn{1}{c}{$N$}                     & \multicolumn{1}{c}{The number of classes in each task}    \\
\multicolumn{1}{c}{$k_{spt}$}                     & \multicolumn{1}{c}{The number of samples of the support set in each task} \\
\multicolumn{1}{c}{$k_{qry}$}                     & \multicolumn{1}{c}{The number of samples of the query set in each task}\\
\multicolumn{1}{c}{$D_{\tau_{i}}^{s}$}                     & \multicolumn{1}{c}{Support set of localization task $\tau_{i}$ that contains $k_{spt}$ number of samples under each location of $N$ ways} \\
\multicolumn{1}{c}{$D_{\tau_{i}}^{q}$}                     &
\multicolumn{1}{c}{Query set of localization task $\tau_{i}$ that contains $k_{qry}$ number of samples under each location of $N$ ways} \\
\multicolumn{1}{c}{$\mathcal{L}_{\tau_{i}}(f_{\bm{\theta}},D_{\tau_{i}}^{s})$}                     & \multicolumn{1}{c}{The task-specific loss function for task $\tau_{i}$ based on model parameters $\bm{\theta}$ and support set $D_{\tau_{i}}^{s}$} \\
\multicolumn{1}{c}{$\mathcal{L}_{\tau_{i}}(f_{\bm{\theta}},D_{\tau_{i}}^{q})$}                     & \multicolumn{1}{c}{The task-specific loss function for task $\tau_{i}$ based on model parameters $\bm{\theta}$ and query set $D_{\tau_{i}}^{q}$} \\
\multicolumn{1}{c}{$\bm{\theta}_{i}^{\prime}$}                     & \multicolumn{1}{c}{The task-specific parameters after the inner loop via one step of gradient descent} \\
\multicolumn{1}{c}{$\bm{\theta}^*$}                     & \multicolumn{1}{c}{Optimal meta-parameters after the outer loop} \\
\multicolumn{1}{c}{$\bm{\theta}_{T}(Q)$}                     & \multicolumn{1}{c}{The task-specific adapted model parameters obtained by updating $\bm{\theta}^*$ after $Q$ steps of gradient descent} \\
\multicolumn{1}{c}{$\alpha$}                     & \multicolumn{1}{c}{The step size of the inner loop}          \\
\multicolumn{1}{c}{$\beta$}                     & \multicolumn{1}{c}{The step size of the outer loop}          \\
\bottomrule[1.5pt] 
\end{tabular}
\label{tab:notations}
\end{table*}

\subsection{Contributions}
The paper introduces a pioneering localization framework called \emph{MetaLoc}, which leverages the power of meta-learning to improve fingerprinting-based indoor localization. It comprises two paradigms: (1) a centralized paradigm based on the vanilla model-agnostic meta-learning (MAML)~\cite{finn2017model} and (2) a distributed paradigm combining MAML with task similarity (MAML-TS) and MAML with domain generalization (MAML-DG). MetaLoc distinguishes itself from existing methods by quickly and efficiently adapting to new environments through a small number of newly collected measurements. The  framework uses the meta-parameters learned from historical environments to initialize a neural network, breaking through the traditional environment-specific localization bottleneck.
\begin{figure*}[h]
  \centering
  \subfigure[CSI fingerprints for different channels at location 1.]{
  \includegraphics[scale=0.8]{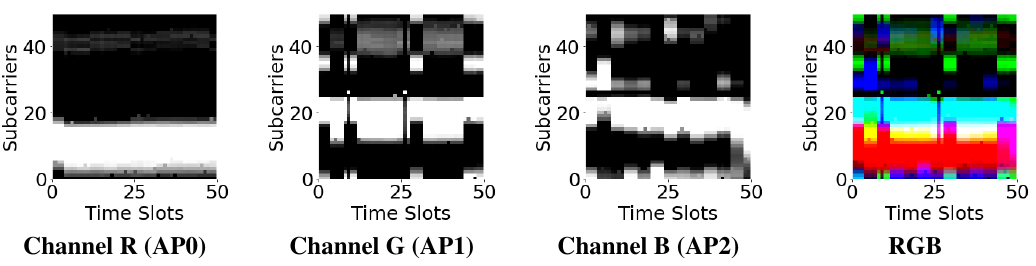}}
  \subfigure[CSI fingerprints for different channels at location 2.]{
  \includegraphics[scale=0.8]{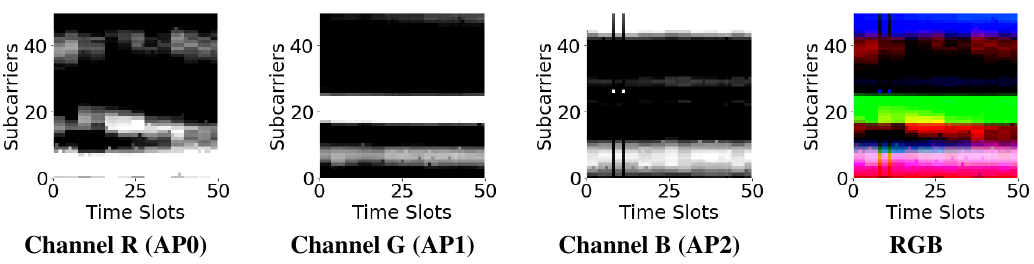}}
  \caption{The newly designed CSI fingerprints.}
  \label{fig:csi-ff}
\end{figure*}
This work expands upon our conference paper ~\cite{gao2022metaloc} with the following aspects. Firstly, we introduce a new algorithm called MAML-DG, which utilizes a dual-component loss function to improve the learning efficiency, thereby offering an advancement over the vanilla MAML algorithm employed in our conference paper. Secondly, we enrich our validation through additional real-world datasets, thus offering a more comprehensive evaluation of our framework.  Lastly, we incorporate insights pertinent to emergent 6G scenarios, suggesting that the philosophy of learning-to-learn has implications for wireless localization, notably in the next-generation intelligent networks.

Overall, the main contributions of this work are fourfold:
\begin{enumerate}
\item \textbf{Outstanding Localization Performance}: MetaLoc is the first work to harness the potential of meta-learning for wireless localization. Its ability to quickly adapt to environmental changes with computationally inexpensive updates sets it apart from the competition. Furthermore, MetaLoc achieves this adaptation with just three CSI images per point in a new environment, solidifying its position as a cost-effective solution.
\item \textbf{Comprehensive Real-World Dataset:} To fairly evaluate the performance of different localization methods, we have created a publicly available dataset. Our dataset was built using a smartphone as the receiver and three different types of WiFi routers as the transmitters. We conducted field trials on two different scenarios: a spacious hall with minimal obstacles and a cluttered lab. To ensure comprehensive data collection, we took five independent measurements over different days, each at 90 grid points. This resulted in 260 CSI images and corresponding RSS vectors per grid point, providing a rich source of data for comparison.
\item \textbf{Innovative  Proposed Paradigms}: We propose two paradigms for training meta-parameters in wireless localization. Our first paradigm, a centralized approach, trains the meta-parameters using all historical environments, resulting in a simplified implementation process. Our second paradigm, a distributed approach, protects the privacy of each environment by training environment-specific meta-parameters. The extensive theoretical analyses and experimental results provide evidence of the superiority of our innovative paradigms.
\item \textbf{Flexible Model Compatibility}: 
The work sets its sights on indoor fingerprinting-based localization that leverages both RSS and CSI wireless signal features. Essentially, MetaLoc can be utilized with any model that has been trained through gradient descent, and any wireless signal features that can be input into a learning model. This compatibility makes it a highly valuable tool for solving various data-driven wireless localization challenges, including both regression and classification problems, and holds tremendous potential for the future of wireless localization.
\end{enumerate}
The remainder of the paper is organized as follows: Section II gives the preliminaries of indoor localization. The proposed MetaLoc is then presented in detail in Section III. Section IV outlines the experimental setup. In Section V, we demonstrate the efficacy of the proposed scheme through computer simulations and real-world data, while future work and challenges towards 6G are presented in Section VI. Finally, the paper is concluded in Section VII. For clarity, the notations adopted throughout the paper are summarized
in Table~\ref{tab:notations}.
\section{Preliminaries}
To validate the proposed framework, we focus on the two widely-used wireless signal features for indoor localization: RSS and CSI. A comprehensive overview of fingerprinting-based localization is also provided in this section.
\subsection{Received Signal Strength (RSS)}
RSS is a metric in decibels that measures the strength of the radio signal during propagation. It can be easily obtained in various real-world wireless networks without the need for additional infrastructure. The widely adopted path-loss model captures the signal attenuation in indoor environments at the GHz frequency band as follows~\cite{bose2007practical}:
\begin{equation}
P_{t} - P_{r} = 10n\lg (d)-G+20\lg\left(\frac{4\pi}{\lambda}\right)+X_{\sigma},
\end{equation}
where $P_{t}$ represents the transmit power of the APs, and $P_{r}$ is the RSS at a RP that is located $d$ meters away from the transmitter. The path-loss exponent $n$ accounts for the impact of obstacles such as walls and doors, whose value is typically greater than two in buildings with blocked paths. $G$ represents the antenna gain, and $\lambda$ is the wavelength of the wireless signal. $X_{\alpha}$ is a normal random variable with a standard deviation (std) of $\sigma$, which ranges from 3 dB to 20 dB.

Although the vanilla path-loss model accounts for the impact of distance and environmental factors on signal strength, it often falls short in accurately describing complex signal attenuation due to the crude nature of the RSS information, making it susceptible to variability in real-world environments.
\subsection{Channel State Information (CSI)}
Orthogonal frequency division multiplexing (OFDM) is a widely used technology in the latest wireless communication standards, including 802.11a, 802.11n, and 802.11ac. The CSI extracted from the OFDM receivers can reveal the multipath characteristics of the wireless channel. In general, CSI is defined as the ratio of received signal to transmitted signal and is represented by a complex number, $H_{i}=\left|H_{i}\right| e^{j \sin \left(\angle H_{i}\right)}$, where $\left|H_{i}\right|$ and $\angle H_{i}$ represent the amplitude and phase of the $i$-th subcarrier, respectively. The Intel 5300 CSI tool~\cite{halperin2011tool}, a commonly used CSI collection toolkit, requires extensive hardware support and a successful connection to each AP. In contrast, we employ Nexmon [47], a more accessible smartphone-based CSI collection tool that does not require connection to surrounding encrypted APs.

In conclusion, CSI is a detailed and sophisticated measurement in the frequency domain compared with the convenient RSS. With multiple subcarriers providing information about different fading or scattered paths, CSI allows for the construction of robust fingerprints that characterize each location and enable the design of accurate localization systems.
\subsection{Fingerprinting-based Localization}
\begin{figure}[ht]
\centering
\includegraphics[width=3.3in]{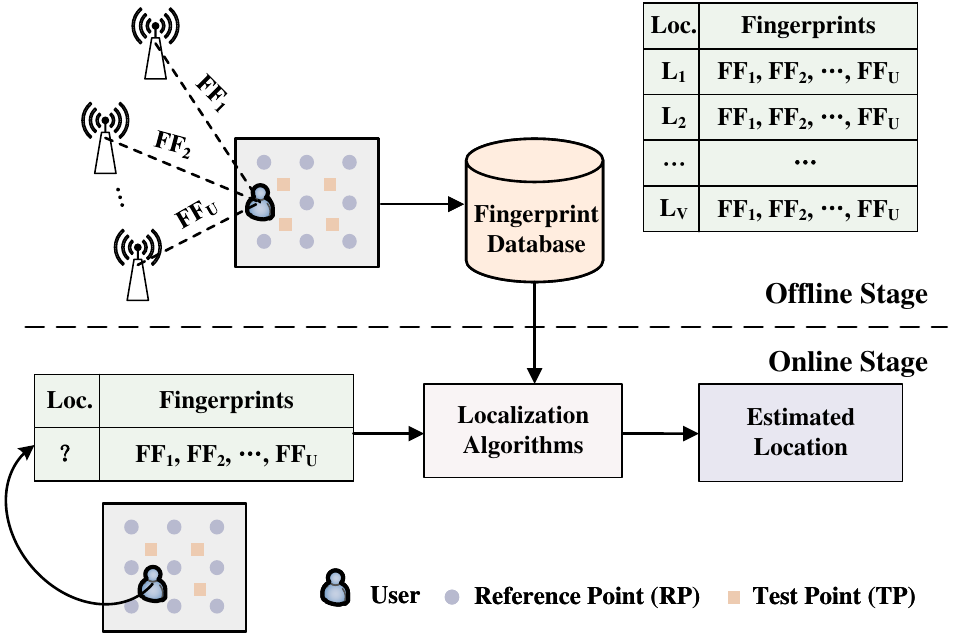}
\caption{A diagram of fingerprinting-based localization, where FF represents the fingerprint fragment collected from each AP, which can be RSS or CSI. Loc. represents the location information.}
\label{fig:fingerprint}
\end{figure}
Figure~\ref{fig:fingerprint} illustrates the diagram of fingerprinting-based localization. The process is divided into two stages: the offline stage and the online stage. In the offline stage, a database includes fingerprints and the corresponding location information is created. This involves deploying multiple APs as transmitters, and a user with a mobile device as the receiver. The signal features collected from each AP, such as received RSS or CSI, form a fingerprint fragment (FF). When multiple FFs are received from $U$ APs, they are combined to create a fingerprint that characterizes a specific location $L_{i}$ ($i=1,2,\dots,V$). In the online stage, fingerprints are collected at TPs to estimate the location using localization algorithms.
\begin{figure}[ht]
\centering
\includegraphics[width=3.6in]{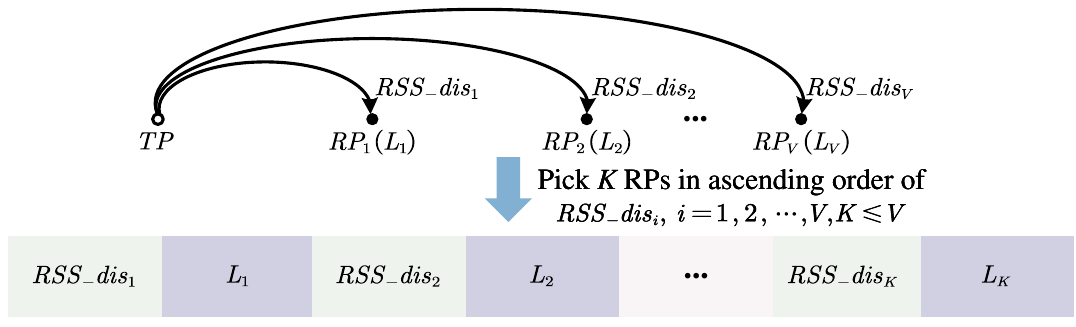}
\caption{The designed RSS fingerprints. }
\label{fig:input_data}
\end{figure}
Next, we will delve into our proposed fingerprint design that uses both RSS and CSI.
\subsubsection{RSS}
To determine the location of each TP in our fingerprint database, we utilize the estimated TP's $K$-closest RPs in the signal space. This information includes the $K$-nearest Euclidean distances and the physical coordinates of the corresponding RPs, as depicted in Fig.~\ref{fig:input_data}. In particular, $RSS\_dis_{i}$ represents the $i$-th closest Euclidean distance between the estimated TP and all the RPs in the signal space, while $L_{i}$ denotes the corresponding location of the $i$-th RP.
\subsubsection{CSI}
Nexmon has the ability to extract 52 valuable subcarriers from each AP, allowing us to create CSI images with amplitude information for each location, as seen in Fig.~\ref{fig:csi-ff}. The CSI images, with dimensions $52\times50\times3$, consist of 50 data packets (rows) and 52 measured amplitude values (columns) from subcarriers. The three channels of the CSI image represent information received from the three APs deployed in the environment, similar to the RGB channels in a colorful image. The CSI images highlight distinctive characteristics between two separate  locations as shown in Fig.~\ref{fig:csi-ff}, making them ideal candidates as fingerprints for localization.

Moreover, we use a histogram intersection method to assess the similarity of two different CSI images. Given an image, we divide all its pixels into equal-interval bins and create a histogram, where each bar represents the pixel count in that bin. Let $Z_{j}(I)$ and $Z_{j}(I')$ denote the pixel count in the $j$-th bin of images $I$ and $I'$, respectively. The histogram intersection of two images each with $n$ bins is defined as ~\cite{hwang2009model}:
\begin{equation}
Z(I) \cap Z\left(I^{\prime}\right)=\sum_{j=1}^n \min \left(Z_j(I), Z_j\left(I^{\prime}\right)\right),\label{eq:HI}
\end{equation}
where $\min\left(x,y\right)$ function takes two values $x$ and $y$ as arguments and returns the smaller one. The histogram intersection calculates the similarity between the two images by summing up the minimum number of pixels in each bin of both images. In other words, it measures the overlap between the two histograms. A larger histogram intersection value indicates that the two images are more similar. In the following experiments, $n=256$ bins are used for the designed CSI images with pixel values ranging from 0 to 255.
\section{The Proposed System}
\begin{figure*}[ht]
\centering
\includegraphics[scale=0.35]{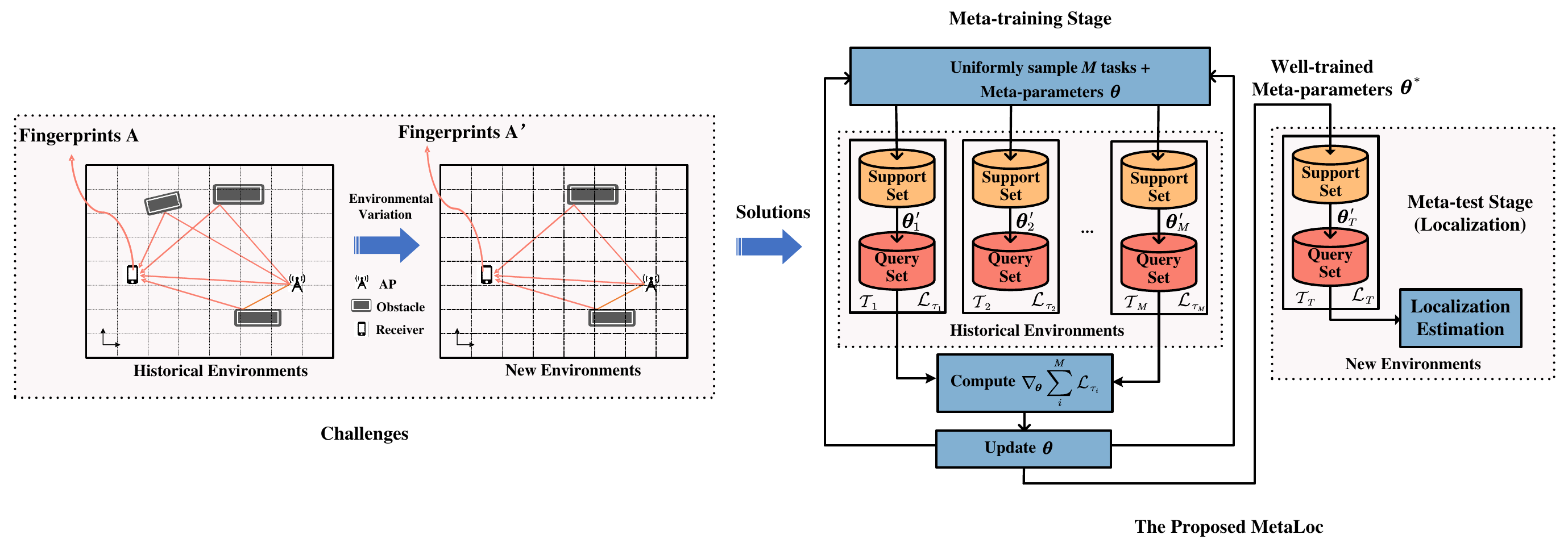}
\caption{The overview of the proposed MetaLoc framework.}
\label{fig:overview}
\end{figure*}
Figure~\ref{fig:overview} presents the challenges and our MetaLoc framework. In transmitter-receiver settings, environmental changes like moving obstacles can significantly alter signal propagation and, consequently, location fingerprints. This may invalidate the existing fingerprint database, necessitating the update for new neural network training. We refer to previous environments where calibration was done as historical environments and to the environments where localization is to be performed as new environments. 

The proposed MetaLoc is divided into two stages: meta-training and meta-test. The meta-training stage requires data gathered from historical environments, while the meta-test stage only necessitates a minimal amount of data in the new  environment. The meta-parameters $\bm{\theta}^{*}$ learned from the meta-training stage provide a strong starting point for the training process in the meta-test stage, eliminating the need to start from scratch in the face of environmental variations.
\subsection{Meta-learning}
Meta-learning is a learning to learn approach that enables the learning model to adapt to new tasks by leveraging previous experience from related tasks. In this framework, tasks are drawn from a specific distribution, denoted as $\tau\sim\mathcal{P}(\tau)$, and each task includes a support set for training and a query set for test. In an $N$-way $k$-shot classification problem, a task consists of $N$ classes, each with $k$ samples. In the meta-training stage, $M$ training tasks, $\{\tau_{i}\}_{i=1}^{M}\sim\mathcal{P}(\tau)$, are sampled from the distribution and the corresponding datasets are made available to the model. In the meta-test stage, a new test task $T\sim\mathcal{P}(\tau)$ is presented, consisting of a small support set and a query set. The objective of meta-learning is to train a model on the $M$ training tasks, such that it can quickly adapt to the new test task using the small support set and perform well on the query set. 

Model-agnostic meta-learning (MAML) achieves this by learning a set of initial parameters $\bm{\theta}_{MAML}$ for neural networks that enable good performance on a new task with only a few gradient descent steps. In the meta-training stage, MAML formulates a meta-optimization problem to find $\bm{\theta}_{MAML}$ as:
\begin{equation}    \bm{\theta}_{MAML}=\mathop{\arg\min}\limits_{\bm{\theta}}\sum_{i=1}^{M}\mathcal{L}_{\tau_{i}}\left(\bm{\theta}-\alpha\nabla_{\bm{\theta}}\hat{\mathcal{L}}_{\tau_{i}}(\bm{\theta})\right),
\end{equation}
where it contains two task-specific loss functions $\hat{\mathcal{L}}_{\tau_{i}}$ and
$\mathcal{L}_{\tau_{i}}$ computed based on the support set and query set of the training task $\tau_{i}$, respectively. Then the meta-parameters are updated via stochastic gradient descent (SGD):
\begin{equation}
\bm{\theta}_{MAML}\gets\bm{\theta}_{MAML}-\beta\nabla_{\bm{\theta}}\sum_{i=1}^{M}\mathcal{L}_{\tau_{i}}\left(\bm{\theta}-\alpha\nabla_{\bm{\theta}}\hat{\mathcal{L}}_{\tau_{i}}(\bm{\theta})\right),
	\label{eq:outer_loop}
\end{equation}
where $\alpha$ and $\beta$ denote the step size of the inner loop and outer loop, respectively. During the meta-test stage, the meta-parameters $\bm{\theta}_{MAML}$ are fine-tuned to obtain the parameters $\bm{\theta}_{T}$ for the neural network used in the test task $T$. This is achieved by updating the meta-parameters using the gradient of the loss function $\hat{\mathcal{L}_{T}}\left(\bm{\theta}_{MAML}\right)$ computed based on the support set of the test task, as follows:
\begin{equation}\bm{\theta}_{T}\gets\bm{\theta}_{MAML}-\alpha\nabla_{\bm{\theta}}\hat{\mathcal{L}_{T}}\left(\bm{\theta}_{MAML}\right),\end{equation} 

In the subsequent sections, we introduce two paradigms for implementing MetaLoc to learn the meta-parameters: the centralized paradigm and the distributed paradigm. The centralized paradigm shares data collected from all historical environments to derive the meta-parameters, while the distributed paradigm maintains the privacy of data in each environment by training environment-specific meta-parameters separately. The advanced distributed paradigm of MetaLoc overcomes the domain shift challenge by leveraging the exchange of environment-specific meta-parameters between different environments.
\subsection{Centralized MetaLoc Paradigm }
The centralized paradigm of MetaLoc is built upon the foundation of the vanilla MAML algorithm~\cite{finn2017model}. In this paradigm, data from various historical environments are collectively employed during the meta-training stage to learn the meta-parameters.
\subsubsection{Meta-training stage}
The primary purpose of the meta-training stage is to derive the well-trained meta-parameters $\bm{\theta}^*$. This is achieved through the use of a neural network that maps the observed fingerprints of the estimated TPs to the desired location outputs. The process is carried out in the following steps: 

\textbf{Step \ding{172}}: The neural network is initialized with a preselected network architecture and a set of randomly initialized meta-parameters $\bm{\theta}$, which are represented by the parameterized function $f_{\bm{\theta}}$. A total of $M$ tasks ${\tau_{1},\dots,\tau_{M}}$ are sampled from the localization task distribution $p(\tau)$ to be used as the training tasks. Each task $\tau_{i}$ is composed of a loss function $\mathcal{L}_{\tau_{i}}$, a support set $D_{\tau_{i}}^{s}$ and a query set $D_{\tau_{i}}^{q}$. The loss function $\mathcal{L}_{\tau_{i}}$ provides task-specific feedback, with cross-entropy being used for classification tasks and either mean-squared-error (MSE) or root-mean-square-error (RMSE) being used for regression tasks. 

\textbf{Step \ding{173}}: For localization task $\tau_{i}$, we train $f_{\bm{\theta}}$ with the support set $D_{\tau_{i}}^{s}$ and derive the task-specific loss $\mathcal{L}_{\tau_{i}}(f_{\bm{\theta}};D_{\tau_{i}}^{s})$.
We obtain task-specific parameters $\bm{\theta}^{\prime}_i$ during the inner loop using one-step gradient descent update, expressed as\begin{equation}
	\bm{\theta}^{\prime}_i=\bm{\theta}-\alpha\nabla_{\bm{\theta}}\mathcal{L}_{\tau_{i}}(f_{\bm{\theta}};D_{\tau_{i}}^{s}),\label{eq:inner_update}
\end{equation}where the hyper-parameter $\alpha$ represents the step size of the inner loop. The task-specific parameters $\bm{\theta}^{\prime}_{i}$ obtained in Step \ding{173} provide only limited information about each localization task, as they are derived using a one-step gradient descent update on the support set $D_{\tau_{i}}^{s}$. Further evaluation of their performance on the query set $D_{\tau_{i}}^{q}$ is necessary to obtain a more comprehensive understanding of each task. 

\textbf{Step \ding{174}}: The performance of one-step update for the $i$-th localization  task can be further evaluated using the samples in the query set given by $\mathcal{L}_{\tau_{i}}(f_{\bm{\theta}^{\prime}_i};D_{\tau_{i}}^{q})$. We define the meta-loss as the sum of all task-specific losses, $\sum_{i=1}^{M}\mathcal{L}_{\tau_{i}}(f_{\bm{\theta}^{\prime}_i};D_{\tau_{i}}^{q})$. This meta-loss is also called the meta-objective.  We then update the meta-parameters $\bm\theta$ by minimizing the meta-loss. In this way, the meta-optimization across all $M$ tasks is given by
\begin{equation}
	\bm{\theta}^{*}
	=\underset{\bm{\theta}}{\arg\min}
	\sum_{i=1}^{M} 
	\mathcal{L}_{\tau_{i}}
	\left(
	f_{\bm{\theta}^{\prime}_i};D_{\tau_{i}}^{q}
	\right).
\end{equation}
It is important to note that during the meta-optimization process, the optimization is performed over the meta-parameters $\bm{\theta}$, while the meta-objective is computed using the task-specific parameters $\bm{\theta}_i^{\prime}$. This separation ensures that the meta-parameters are updated to achieve the best generalization performance across all tasks, while the task-specific parameters are used to evaluate the one-step update performance for each individual task.

\textbf{Step \ding{175}}: The meta-optimization during outer loop is carried out using SGD, where the meta-parameters $\bm{\theta}$ are updated as follows:
\begin{equation}
\bm{\theta}\gets\bm{\theta}-\beta\nabla_{\bm{\theta}}\sum_{i=1}^{M}\mathcal{L}_{\tau_{i}}\left(f_{\bm{\theta}^{\prime}_i};D_{\tau_{i}}^{q}\right),
	\label{eq:outer_loop}
\end{equation}
where $\beta$ is the step size for the outer loop. During this optimization, the meta-objective is computed using the adapted task-specific parameters $\bm{\theta}_i^{\prime}$. To compute the partial derivative of $\mathcal{L}_{\tau_{i}}(f_{\bm{\theta}^{\prime}_i};D_{\tau_{i}}^{q})$ with respect to the $m$-th element of $\bm{\theta}$, we can use the chain rule and write:
\begin{equation}
	\frac{\partial\mathcal{L}_{\tau_{i}}(f_{\bm{\theta}^{\prime}_i};D_{\tau_{i}}^{q})}
	{\partial\boldsymbol{\theta}^{(m)}}
	=\sum_{n}
	\frac{\partial\mathcal{L}_{\tau_{i}}\left(f_{\bm{\theta}^{\prime}_i};D_{\tau_{i}}^{q}\right)}
	{\partial\bm{\theta}_i^{\prime(n)}}
	\frac
	{\partial\bm{\theta}_i^{\prime(n)}}
	{\partial\bm{\theta}^{(m)}}
	\label{eq:partial1},
\end{equation}
and by merging Eq.~(\ref{eq:inner_update}), we have
\begin{equation}
	\frac
	{\partial\bm{\theta}^{\prime(n)}_i}
	{\partial\boldsymbol{\theta}^{(m)}} 
	=
	\frac{\partial\bm{\theta}^{(n)}}{\partial\boldsymbol{\theta}^{(m)}}-\alpha\frac{\partial\mathcal{L}_{\tau_{i}}(f_{\boldsymbol{\theta}};D_{\tau_{i}}^{s})}{\partial\boldsymbol{\theta}^{(n)}\partial\boldsymbol{\theta}^{(m)}}\label{eq:partial2}.
\end{equation}
However, directly computing the partial derivative $\frac{\partial\bm{\theta}_i^{\prime(n)}}{\partial\boldsymbol{\theta}^{(m)}}$ using the inner loop update rule involves computing second-order derivatives, which can be computationally expensive. Therefore, we make use of the first-order approximation~\cite{goodfellow2014explaining}, assuming that the second-order derivatives are negligible. Specifically, we approximate $\frac{\partial\bm{\theta}_i^{\prime(n)}}{\partial\boldsymbol{\theta}^{(m)}}$ as follows:\\
1) When $m\neq n:$
\begin{equation}
	\frac{\partial\bm{\theta}_i^{\prime(n)}}{\partial\boldsymbol{\theta}^{(m)}}  =-\alpha\frac{\partial\mathcal{L}_{\tau_{i}}(f_{\boldsymbol{\theta}};D_{i}^{s})}{\partial\boldsymbol{\theta}^{(n)}\partial\boldsymbol{\theta}^{(m)}} \approx0\label{eq:mneqn}.
\end{equation}
2) When $m=n:$
\begin{equation}
\frac{\partial\bm{\theta}_i^{\prime(n)}}{\partial\boldsymbol{\theta}^{(m)}}=1-\alpha\frac{\partial\mathcal{L}_{\tau_{i}}(f_{\boldsymbol{\theta}};D_{\tau_{i}}^{s})}{\partial\boldsymbol{\theta}^{(n)}\partial\boldsymbol{\theta}^{(m)}}\approx1. \label{eq:meqn}
\end{equation}
Using these approximations, Eq.~(\ref{eq:partial1}) can be simplified to:
\begin{equation}
\frac{\partial\mathcal{L}_{\tau_{i}}(f_{\boldsymbol{\theta}^{\prime}_i};D_{\tau_{i}}^{q})}{\partial\boldsymbol{\theta}^{(m)}}\approx\frac{\partial\mathcal{L}_{\tau_{i}}(f_{\boldsymbol{\theta}^{\prime}_{i}};D_{\tau_{i}}^{q})}{\partial\boldsymbol{\theta}^{\prime(m)}_{i}}.
\end{equation}
Therefore, the meta-optimization over the meta-parameters $\bm{\theta}$ can be simplified to optimization over the task-specific parameters $\bm{\theta}_{i}^{\prime}$, which eliminates the need for complicated second-order derivatives.
\subsubsection {Meta-test Stage}
After the meta-training stage, we obtain well-trained meta-parameters $\bm{\theta}^{*}$ that capture the essential characteristics of localization from historical environments. To perform an unseen task in the new environment $T\sim\mathcal{P}(\tau)$ with support set $D_T^{s}$ and query set $D_T^{q}$, we use the pre-trained neural network ${f_{\bm{\theta}^{*}}}$ with a predefined architecture and learned meta-parameters $\bm{\theta}^*$ as initialization. Then, we update task-specific adapted parameters $\bm{\theta}_{T}(Q)$ by taking $Q$-steps of gradient descent on the small support set $D_T^{s}$, which is given by:
\begin{equation}
	\begin{small}
		\begin{gathered}	\bm{\theta}_{T}(Q)\!=\!\bm{\theta}^{*}\!\!-\!\alpha\!\left[\!\nabla_{\bm{\theta}^{*}} \mathcal{L}_{T}\left(f_{\bm{\theta}^{*}}\text{\!;}D_{T}^{s}\right)\!+\!\sum_{j=1}^{Q-1}\! \nabla_{\bm{\theta}_{T}(j)} \mathcal{L}_{T}\!\left(f_{\bm{\theta}_{T}\left(j\right)}\text{\!;}D_{T}^{s}\right)\!\right],
			\end{gathered}
	\end{small} 
\end{equation}
and task-specific optimal parameters $\bm{\theta}_{T}^{*}$ for test task $T$ is given by \begin{equation}\begin{gathered}	{\bm{\theta}_{T}^{*}=\underset{\bm{\theta}_{T}\left(Q\right)}{\arg\min}}\mathcal{L}_{T}\left(f_{\bm{\theta}_{T}\left(Q\right)}\text{;}D_{T}^{s}\right),\end{gathered}
\end{equation}
where $\mathcal{L}_{T}\left(f_{\bm{\theta}_{T}\left(Q\right)}\text{;}D_{T}^{s}\right)$ denotes the test loss measured on the query set $D_{T}^{s}$. 
 From the process outlined above, it is clear that this procedure exhibits a model-agnostic characteristic. Specifically, it does not depend on a particular structure of the model. Instead, the process is solely dictated by the model's parameters and their gradients, indicating its potential use with any model that utilizes a gradient descent type optimization method for training. For clarity, the entire process is summarized in Algorithm \ref{alg:alg1}.

\begin{algorithm}[ht]
\caption{Vanilla MAML}\label{alg:alg1}
\begin{algorithmic}
\STATE \textbf{Require:} \\$\mathcal{P}(\tau)$: distribution over tasks;
\\$\alpha$: step size of the inner loop; \\$\beta$: step size of the outer loop;
\STATE \textbf{Meta-training Stage (in the historical environments):}
\STATE 1: Randomly initialize $\bm{\bm{\theta}}$;
\STATE 2: For $ite$ in iterations do:
\STATE 3: \hspace{0.5cm}Sample training tasks $\{\tau_{i}\}_{i=1}^{M}\sim\mathcal{P}(\tau) $;
\STATE 4: \hspace{0.5cm}For each $i$ in $\{1,2,\dots,M\}$ do:
\STATE 5: \hspace{1.0cm}$ \bm{\bm{\theta}}_{i}^{\prime}=\bm{\bm{\theta}} - \alpha\nabla_{\bm{\bm{\theta}}}\mathcal{L}_{\tau_{i}}(f_{\bm{\theta}};D_{\tau_{i}}^{s})$;
\STATE 6: \hspace{0.5cm}$ \bm{\theta}\gets\bm{\theta}-\beta\nabla_{\bm{\theta}}\sum_{\tau_{i}}\mathcal{L}_{\tau_{i}}(f_{\bm{\theta}_{i}^{\prime}};D_{\tau_{i}}^{q})$;
\STATE 7: \textbf{return}  $\bm{\theta}^*\gets\bm{\theta}$ when it converges.
\STATE \textbf{Meta-test Stage (in the new environment):}
\STATE 8: Sample a test task $T\sim\mathcal{P}(\tau) $; 
\STATE 9: $\bm{\theta}_{T}\gets\bm{\theta}^*-\alpha\nabla_{\bm{\theta}}\mathcal{L}_{T}(f_{\bm{\theta}^*};D_{T}^{s})$;
\STATE 10: \textbf{return}  $\bm{\theta}_{T}^*\gets\bm{\theta}_T$ when it converges.
\end{algorithmic}

\end{algorithm}
\subsubsection{Performance analysis of the vanilla MAML}

While Finn \textit{et al.}\cite{finn2017model} demonstrated the effectiveness of the vanilla MAML in various tasks, they did not provide a comprehensive theoretical analysis to support its effectiveness using few-gradient based adaptation. Later, Zhou \textit{et al.} provided performance guarantees and some deeper insights into the superiority of MAML over traditional neural network training~\cite{zhou2021task}. We borrow similar ideas of reference~\cite{zhou2021task} to address the following questions: (1) what is the reason behind the superior performance of the vanilla MAML over traditionally direct training a neural network with a variety of different environments? (2) which factors influence the test performance of the vanilla MAML? To address the aforementioned questions, the loss function of the vanilla MAML is assumed to conform to Definition 1, which sets the stage for the introduction of Theorem 1. The theorem sets a theoretical upper bound on the excess risk incurred during MAML's adaptation to new tasks, providing a quantitative measure of MAML's adaptability. Given the length constraints of our paper, we have chosen not to include a detailed proof of Theorem 1. Interested readers may refer to the comprehensive proof provided in the supplementary materials in \cite{zhou2021task}. 

\noindent\textbf{Definition 1.} (Lipschitz Continuity). \textit{A function $f(\bm{\theta})$ is said to be G-Lipschitz continuous over a region D (bounded or unbounded) if there exists a G\textgreater0 such that $||f(\bm{\theta}_{1})-f(\bm{\theta}_{2})||_{2}\leq G||\bm{\theta}_{1}-\bm{\theta}_{2}||_{2}$ for all $\bm{\theta}_{1},\bm{\theta}_{2}\in D$. Moreover, $f(\bm{\theta})$ is said to be $W$-smooth if $||\nabla f(\bm{\theta}_{1})-\nabla f(\bm{\theta}_{2})||_{2}\leq W||\bm{\theta}_{1}-\bm{\theta}_{2}||_{2}$}.

Next, the excess risk is defined with respect to $\bm{\theta}_{T}(Q)$ as $ER(\bm{\theta}_{T}(Q)) = E_{T\sim\tau}E_{D_{T}}[\mathcal{L}(\bm{\theta}_{T}(Q))-\mathcal{L}(\bm{\theta}_{T}^*)]$, which evaluates the loss difference
 on all samples from all tasks and well measures the test performance of the adapted parameters $\bm{\theta}_{T}(Q)$ over $Q$ gradient
steps. The lower the $ER(\bm{\theta}_{T}(Q))$, the better adaptation ability the MAML possesses.

\noindent\textbf{Theorem 1.} \textit{Suppose $\mathcal{L}_{T}(f_{\bm{\theta}},D_{T}^s)$ is G-Lipschitz continuous and $W$-smooth with respect to the parameters $\bm{\theta}$, and $\alpha$ satisfies $\alpha\leq\frac{1}{W}$. Setting $\rho = 1+2\alpha W$, then for any $T\sim \mathcal{P}(\tau)$ with $D_T^{s}=\left\{\left(x_i, y_i\right)\right\}_{i=1}^{k_{spt}} \sim T$, we have} 
\begin{small}\begin{align*}
  ER(\bm{\theta}_{T}^{Q}) &\leq \frac{2G^{2}(\rho^{Q}-1)}{k_{spt}*W}+E_{T\sim\tau}E_{D_{T}^s}[\mathcal{L}_{T}(\bm{\theta}_{T}^{Q};D_{T}^s)-\mathcal{L}_{T}(\bm{\theta}_{T}^*)] \\
    &\leq \frac{2G^{2}(\rho^{Q}-1)}{k_{spt}*W}+\frac{1}{2\alpha}E_{T\sim\tau}[||\bm{\theta}^*-\bm{\theta}_{T}^*||_{2}^{2}].
\end{align*}\end{small}

The second inequality implies that a smaller expected distance between $\bm{\theta}^*$ and $\bm{\theta}_{T}^*$ over $T$ (i.e., $E_{T\sim\tau}[||\bm{\theta}^*-\bm{\theta}_{T}^*||_{2}^{2}]$) leads to smaller $ER(\bm{\theta}_{T}(Q))$. 

\begin{figure}
\centering
\includegraphics[width=3in,angle=0]{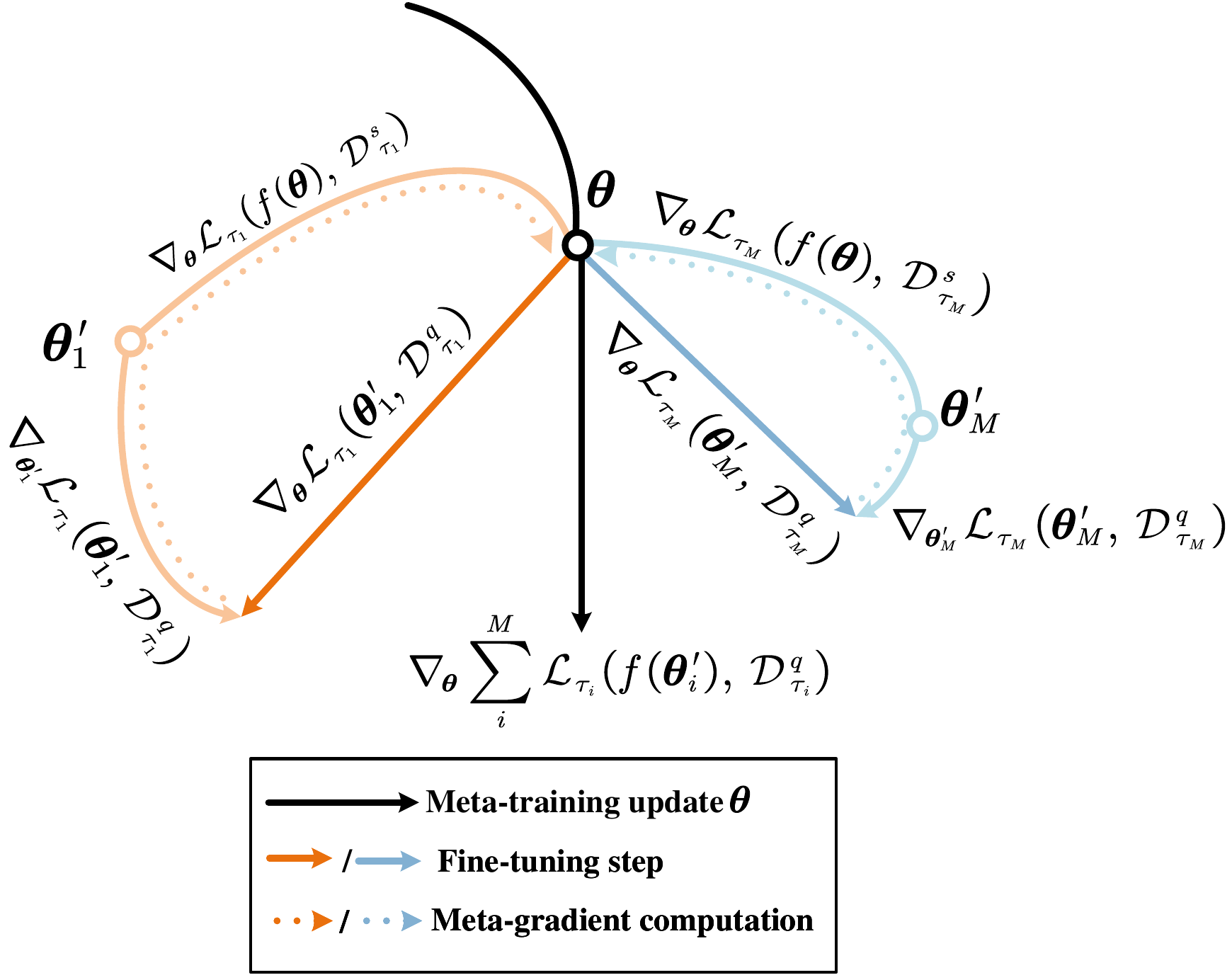}
\label{fig:maml}
\caption{Diagram of MAML, which optimizes the meta-parameters $\bm{\theta}$ to obtain the fast adaptation to new tasks.}
\label{fig:par}
\end{figure}

Following the above idea, we compare the vanilla MAML with a conventional neural network in the next. The vanilla MAML trains the meta-parameters $\bm{\theta}$ as shown in Fig.~\ref{fig:par}, which demonstrates the paths in the parameters space with $M$ tasks. An inner loop is first conducted based on the support set of each task and obtains the task-specific parameters $\bm{\theta}_{i}^{\prime},i=1,\dots,M$. Next, an outer loop is implemented to find each task's most potential direction toward the optimal parameters based on the query set. Finally, we determine the direction towards the optimal parameters for each training task (represented in different colors). The meta-parameters $\bm{\theta}$ are then updated based on the average across the optimal directions of these training tasks (path shown in black). Through this approach, MAML updates the meta-parameters $\bm{\theta}$ in a direction that aligns with all training tasks, with each $\bm{\theta}_{i}^*$ receiving equal weight in the gradient descents of the outer loop. Consequently, it is expected that $E_{T\sim\tau}[||\bm{\theta}^*-\bm{\theta}_{T}^*||_{2}^{2}]$ is smaller in MAML than in other traditional training methods.

It should be noted that conventional neural networks do not employ the terminologies of tasks, inner loops, and outer loops. Instead, models are trained using data from a particular dataset at one time. The issue with this approach is that the optimizer may overfit a single environment by finding a path that achieves rapid loss reduction for that environment but shows slow convergence for other environments. In such cases, $E_{T\sim\tau}[||\bm{\theta}^*-\bm{\theta}_{T}^*||_{2}^{2}]$ would be larger than in MAML, as demonstrated in Section V.A.

\begin{algorithm}[ht]
\caption{MAML-TS}\label{alg:TS}
\begin{algorithmic}
\STATE \textbf{Require:} \\$\{\mathcal{P}^{(i)}(\tau)\}_{i=1}^{S}$: distributions over tasks in $S$ domains;
\\$\alpha$: step size of the inner loop; \\$\beta$: step size of the outer loop;
\STATE \textbf{Meta-training Stage (in the historical environments):}
\STATE 1: Randomly initialize $\bm{\bm{\theta}}$;
\STATE 2: For environment $i$ in $\{1,2,\dots,S\}$ do:
\STATE 3: \hspace{0.5cm}For $ite$ in iterations do:
\STATE 4: \hspace{1cm}Sample training tasks $\{\tau_{j}\}_{j=1}^{M}\sim\mathcal{P}^{(i)}(\tau)$;
\STATE 5: \hspace{1cm}For each $j$ in $\{1,2,\dots,M\}$ do:
\STATE 6: \hspace{1.5cm}$ \bm{\bm{\theta}}_{j}^{\prime}=\bm{\bm{\theta}} - \alpha\nabla_{\bm{\bm{\theta}}}\mathcal{L}_{\tau_{j}}(f_{\bm{\theta}};D_{\tau_{j}}^{s})$;
\STATE 7: \hspace{0.5cm}$ \bm{\theta}\gets\bm{\theta}-\beta\nabla_{\bm{\theta}}\sum_{\tau_{j}}\mathcal{L}_{\tau_{j}}(f_{\bm{\theta}_{j}^{\prime}};D_{\tau_{j}}^{q})$;
\STATE 8: \hspace{0.5cm}\textbf{return}  $\bm{\theta}_{i}^*\gets\bm{\theta}$ when it converges.
\STATE \textbf{Meta-test Stage (in the new environment):}
\STATE 9: Sample a test task $T\sim\mathcal{P}(\tau) $; 
\STATE 10: Compute $\text{MMD}_{i}$ between the test task and training tasks from each historical environment $\{\tau_{j}\}_{j=1}^{M}\sim\mathcal{P}^{(i)}(\tau), i=1,2,\dots,S$; 
\STATE 11: Choose the optimal environment-specific meta-parameters $\bm{\theta}_{i^{*}}^*$, where $i^{*} = \arg\min_{i \in \{1, ..., S\}} \text{MMD}_i$.
\STATE 12: $\bm{\theta}_{T}\gets\bm{\theta}_{i^*}^*-\alpha\nabla_{\bm{\theta}}\mathcal{L}_{T}(f_{\bm{\theta}^*};D_{T}^{s})$;
\STATE 13: \textbf{return}  $\bm{\theta}_{T}^*\gets\bm{\theta}_T$ when it converges.
\end{algorithmic}
\end{algorithm}
\subsection{Distributed MetaLoc Paradigm}
 Compared to the centralized paradigm, which is trained on all historical environments, our conference work~\cite{gao2022metaloc} proposes an environment-specific meta-parameter approach. This involves clustering the historical environments based on their propagation models and training a unique set of meta-parameters for each group. Instead of using the traditional meta-parameters trained on all tasks, we select the best environment-specific meta-parameters for the target localization task based on task similarity between the test task and the historical environments. To measure this task similarity concretely, we utilize the maximum mean discrepancy (MMD) measure~\cite{gretton2012kernel}, as shown in
\begin{equation}\begin{gathered}\operatorname{MMD}[\mathcal{G}, \tau_{1}, \tau_{2}]:=\sup _{h \in \mathcal{H}}\left(\mathbf{E}_{x}[h(x)]-\mathbf{E}_{y}[h(y)]\right),
\label{eq:mmd}
\end{gathered}
\end{equation}
where $x$ and $y$ are wireless signal features, such as RSS fingerprints or CSI fingerprints, for a pair of localization tasks $\tau_{1}$ and $\tau_{2}$ in the tasks distribution $\mathcal{P}(\tau)$, respectively. Moreover, $\mathcal{H}$ refers to a class of functions $h$. In \cite{gretton2012kernel}, it was shown that when $\mathcal{H}$ is a unit ball in a universal reproducing kernel Hilbert space defined on $\mathcal{P}(\tau)$ with the associated continuous kernel, then $\operatorname{MMD}[\mathcal{H}, \tau_{1}, \tau_{2}]=0$ if and only if $\tau_{1}$ is equivalent to $\tau_{2}$. A higher MMD value indicates a larger difference between the two tasks.
Given that the algorithm takes into account task similarity, we term it as the model-agnostic meta-learning with task similarity (MAML-TS). The algorithm outlined in Algorithm~\ref{alg:TS}, allows training in isolation for each environment, thereby ensuring better data security. 

To further improve the learning efficiency, we propose model-agnostic meta-learning with domain generalization (MAML-DG). Unlike the MAML-TS approach which trains environment-specific meta-parameters in isolation for each environment, MAML-DG allows the environments to share the environment-specific meta-parameters, while still protecting the raw data of each environment. This is inspired by the idea of imitating real-time train-test domain shifts~\cite{li2018learning} to enable the model to quickly generalize to different domains. In the indoor localization setting, we treat each environment as an individual domain, with the historical environments serving as the training domains and the new environment as the test domain. 
\begin{algorithm}
\caption{MAML-DG}\label{alg:alg2}
\begin{algorithmic}
\STATE \textbf{Require:} \\$\{\mathcal{P}^{(i)}(\tau)\}_{i=1}^{S}$: distributions over tasks in $S$ domains; \\$\alpha$: step size of the inner loop; \\$\beta$: step size of the outer loop; \\$w$: weight of the loss function of the second training domain $D_{II}$;
\STATE \textbf{Meta-training Stage (in the historical environments):}
\STATE 1: Randomly initialize $\bm{\theta}$;
\STATE 2: For $ite$ in iterations do:
\STATE 3: \hspace{0.5cm}Sample two training domains $D_{I}$ and $D_{II}$ uniformly from $\{1,2,...,S\}$;
\STATE 4: \hspace{0.5cm}Sample tasks $\{\tau_{i}^{(D_{I})}\}_{i=1}^{M}\sim\mathcal{P}^{(D_{I})}(\tau) $ in domain $D_{I}$;
\STATE 5: \hspace{0.5cm}For $i$ in range ($M$) do:
\STATE 6: \hspace{1.0cm}$ \bm{\theta}_{i}^{(D_I)}=\bm{\theta} - \alpha\nabla_{\bm{\theta}}\mathcal{L}_{\tau_{i}^{(D_I)}}(f_{\bm{\theta}};D_{\tau_{i}^{(I)}}^{s})$;
\STATE 7: \hspace{0.5cm}$ \bm{\theta}^{\prime}=\bm{\theta}-\beta\nabla_{\bm{\theta}}\sum_{\tau_{i}^{(D_I)}}\mathcal{L}_{\tau_{i}^{(D_I)}}(f_{\bm{\theta}_{i}^{(D_I)}};D_{\tau_{i}^{(D_I)}}^{q}) $;
\STATE 8: \hspace{0.5cm}Sample tasks $\{\tau_{j}^{(D_{II})}\}_{j=1}^{M}\sim\mathcal{P}^{(D_{II})}(\tau)$ in $D_{II}$;
\STATE 9: \hspace{0.5cm}For $j$ in range ($M$) do:
\STATE 10: \hspace{1.0cm}$ \bm{\theta}_{j}^{(D_{II})}=\bm{\theta} - \alpha\nabla_{\bm{\theta}}\mathcal{L}_{\tau_{j}^{(D_{II})}}(f_{\bm{\theta}^{\prime}};D_{\tau_{j}^{(D_{II})}}^{s})$;
\STATE 11:\hspace{0.5cm}$\bm{\theta}\gets\bm{\theta}^{\prime}-w\beta\nabla_{\bm{\theta}}\sum_{\tau_{j}^{(D_{II})}}\mathcal{L}_{\tau_{j}^{(D_{II})}}(f_{\bm{\theta}_{j}^{(D_{II})}};D_{\tau_{j}^{(D_{II})}}^{q}) $;
\STATE 12: \textbf{return}  $\bm{\theta}^*\gets\bm{\theta}$ when it converges.
\STATE \textbf{Meta-test Stage (in the new environment):}
\STATE 13: Sample a test task $T\sim\mathcal{P}(\tau) $; 
\STATE 14: $\bm{\theta}_{T}\gets\bm{\theta}^*-\alpha\nabla_{\bm{\theta}}\mathcal{L}_{T}(f_{\bm{\theta}^*       
      };D_{T}^{s})$;
\STATE 15: \textbf{return}  $\bm{\theta}_{T}^*\gets\bm{\theta}_T$ when it converges.
\end{algorithmic}
\end{algorithm}

As outlined in Algorithm~\ref{alg:alg2}, MAML-DG is designed to train a deep learning model with  parameters $\bm{\theta}$ across $S$ training domains, which may have different statistical distributions but share the same label  and input features space. During each meta-training iteration, MAML-DG randomly selects two training domains $D_{I}, D_{I}=1,2,\dots,S$ and $D_{II}, D_{II}=1,2,\dots,S, D_{I}\neq D_{II}$ and generates tasks in these two domains. The full steps are as follows. 

\textbf{Step \ding{172}}: We virtually train a domain-specific meta-parameters $\bm{\theta}^{\prime}$ on the tasks generated from the training domain $D_{I}$ using the vanilla MAML algorithm. We derive the first domain-specific loss function as
\begin{equation} 
\small
F(\cdot)=\sum_{\tau_{i}^{(D_{I})}}\mathcal{L}_{\tau_{i}^{(D_{I})}}\left(f_{\bm{\theta}_{i}^{(D_{I})}}\right)=\sum_{i=1}^{M}\mathcal{L}_{\tau_{i}^{(D_{I})}}\left(f_{\bm{\theta}-\alpha\nabla_{\bm{\theta}}\mathcal{L}_{\tau_{i}^{(D_{I})}}}\right). 
\end{equation}

\textbf{Step \ding{173}}: With the initialization $\bm{\theta}^{\prime}$ obtained in the previous step, we derive a second domain-specific loss function for the tasks generated from the training domain $D_{II}$ using vanilla MAML once again, which is shown as \begin{equation} 
\small
G(\cdot)=\sum_{\tau_{j}^{(D_{II})}}\mathcal{L}_{\tau_{j}^{(D_{II})}}(f_{\bm{\theta}_{j}^{(D_{II})}})=\sum_{j=1}^{M}\mathcal{L}_{\tau_{j}^{(D_{II})}}\left(f_{\bm{\theta}^{\prime}-\alpha\nabla_{\bm{\theta}^{\prime}}\mathcal{L}_{\tau_{j}^{(D_{II})}}}\right).
\end{equation}

\textbf{Step \ding{174}}: We sum up the two domain-specific losses $F(\cdot)$ and $G(\cdot)$, and then update the meta-parameters $\bm{\theta}$, which emulates the real-time train-test domain shifts and helps the model generalize faster after a few iterations.

These steps are repeated iteratively by randomly sampling from the two different training domains. We provide a performance analysis below to better understand how MAML-DG works. The objective function of MAML-DG is
\begin{equation}
\begin{aligned}
  \mathcal{L}(\bm{\theta}) &= F(\bm{\theta})+w G(\bm{\theta}^{\prime}) \\
    &= F\left(\bm{\theta}\right)+w G\left(\bm{\theta}-\beta\nabla_{\bm{\theta}}\sum_{\tau_{i}^{(D_{I})}}\mathcal{L}_{\tau_{i}^{(D_{I})}}\left(f_{\bm{\theta}_{i}^{(D_{I})}}\right)\right) \\
    &= F(\bm{\theta})+w G\left(\bm{\theta}-\beta\nabla_{\bm{\theta}}F\left(\bm{\theta}\right)\right),
\end{aligned}\label{eq:loss-DG}
\end{equation}
where $F(\bm{\theta})$ is the loss function of the first training domain, whereas $G\left(\bm{\theta}^{\prime}\right)$ is that of the second training domain, with $\bm{\theta}^{\prime}=\bm{\theta}-\beta\nabla_{\bm{\theta}}F\left(\bm{\theta}\right)$ serving as its initialization.

Applying the first-order Taylor's expansion, we derive that:
\begin{equation}
\begin{aligned}
  G\left(\bm{\theta}-\beta\nabla_{\bm{\theta}}F\left(\bm{\theta}\right)\right) &= G\left(\bm{\theta}\right)+\nabla_{\bm{\theta}}G\left(\bm{\theta}\right)\cdot\left(-\beta\nabla_{\bm{\theta}}F\left(\bm{\theta}\right)\right) \\
    &= G\left(\bm{\theta}\right)-\beta\left(\nabla_{\bm{\theta}}G(\bm{\theta})\cdot\nabla_{\bm{\theta}}F(\bm{\theta})\right).
\end{aligned}\label{eq:G}
\end{equation}
Note that the remainder of the above Taylor's expansion is:
\begin{equation}
\begin{aligned}
  Rem
  &= \frac{1}{2}\left(\left(-\beta\nabla_{\bm{\theta}}F(\bm{\theta})\right)^{T}\cdot \nabla_{\bm{\theta}}\nabla_{\bm{\theta}}G(\delta)\cdot (-\beta\nabla_{\bm{\theta}}F(\bm{\theta}))^{T}\right),
\end{aligned}
\end{equation}
\noindent where $\delta$ is a number that lies in between $\bm{\theta}$ and $\bm{\theta}-\beta\nabla_{\bm{\theta}}F(\bm{\theta})$. Plugging Eq.~(\ref{eq:G}) into Eq.~(\ref{eq:loss-DG}) yields:
\begin{equation}
\begin{aligned}
  \mathcal{L}(\bm{\theta}) &= F(\bm{\theta})+w G\left(\bm{\theta}-\beta\nabla_{\bm{\theta}}F\left(\bm{\theta}\right)\right) \\
    &=  F(\bm{\theta})+w G(\bm{\theta})-w\beta\left(\nabla_{\bm{\theta}}G(\bm{\theta})\cdot\nabla_{\bm{\theta}}F\left(\bm{\theta}\right)\right). 
\end{aligned}
\end{equation}

The loss function is composed of two parts: (i) $F(\bm{\theta})+w G(\bm{\theta})$, and (ii) $-w\beta(\nabla_{\bm{\theta}}G(\bm{\theta})\cdot\nabla_{\bm{\theta}}F(\bm{\theta}))$. Minimizing this loss function is equivalent to minimizing both (i) and (ii). Part (i) aims to minimize the loss in both training domains, which is intuitive. Part (ii) is equivalent to maximizing the dot product of the gradients of $\nabla_{\bm{\theta}}G(\bm{\theta})$ and $\nabla_{\bm{\theta}}F(\bm{\theta})$. In other words, we aim to maximize $||\nabla_{\bm{\theta}}F(\bm{\theta})||_{2}\cdot||\nabla_{\bm{\theta}}G(\bm{\theta})||_{2}\cdot cos(\delta)$, where $\delta$ represents the angle between $\nabla_{\bm{\theta}}F(\bm{\theta})$ and $\nabla_{\bm{\theta}}G(\bm{\theta})$. Therefore, the dot product will be larger if $\nabla_{\bm{\theta}}F(\bm{\theta})$ and $\nabla_{\bm{\theta}}G(\bm{\theta})$ tend to have a similar direction. In combining aspects (i) and (ii), the optimizer is designed to guide the loss reduction in both training domains towards a similar direction. This strategy contributes to the accelerated convergence of MAML-DG. In comparison, the commonly employed objective function $F(\bm{\theta})+G(\bm{\theta})$ could lead to slower convergence, as it seeks a path that allows for rapid decrease in one domain but slower convergence in the other.
\begin{figure}
    \centering
    \includegraphics[width=1\linewidth]{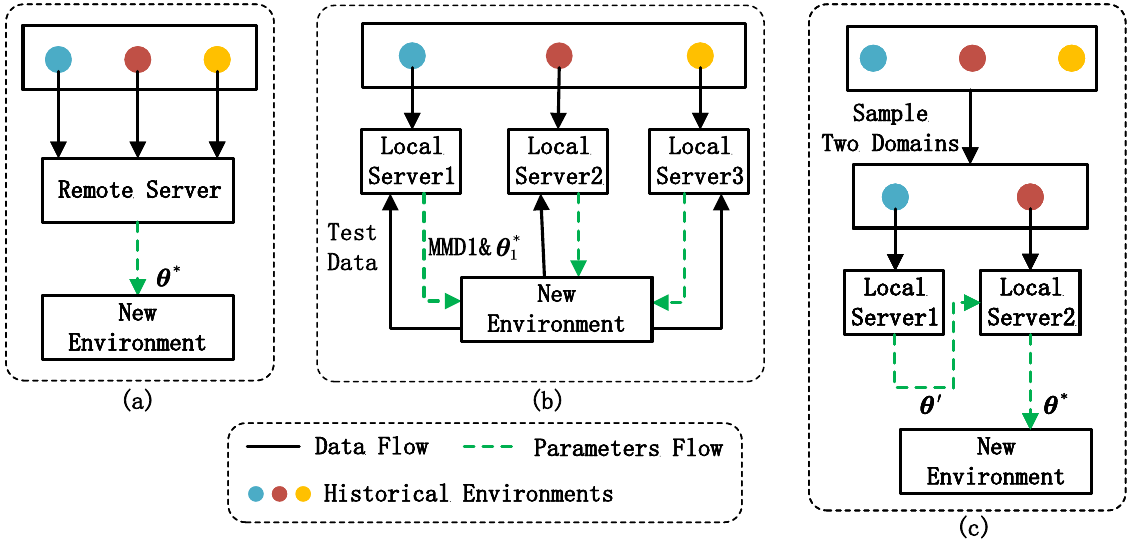}
    \caption{Centralized  paradigm: (a)  vanilla MAML versus distributed paradigm: (b) MAML-TS and (c) MAML-DG.}
    \label{fig:connections}
\end{figure}

\subsection{Comparison between Paradigms}
Figure~\ref{fig:connections} summarizes our proposed paradigms. For clarity, we utilize three distinct environments as examples, each denoted by differently colored circles. Vanilla MAML depicted in Fig.~\ref{fig:connections} (a) relies on a comprehensive data collection process from all environments to derive the meta-parameters $\boldsymbol{\theta}^{*}$. This process necessitates transferring raw wireless signal data, including RSS and CSI, to a remote server, thereby introducing potential privacy concerns. In contrast, the proposed MAML-TS, displayed in Fig.~\ref{fig:connections} (b), calculates environment-specific meta-parameters independently within each local environment. In this method, each historical environment has its own local server. The task-similarity MMD between the test dataset and each historical environment is computed in their own local server, ensuring raw data from the historical environments is unexposed, and data privacy is thus maintained.
As for MAML-DG shown in Fig.~\ref{fig:connections} (c), for each iteration we start the process by randomly sampling two representatives from the pool of all environments. The meta-parameters shared across domains, as indicated in lines 7 and 11 of Algorithm 3, are tied to the neural network model and do not explicitly reveal any user-specific data. Although there are established measures to confront such privacy concerns, a detailed investigation of these strategies is outside the scope of our current research~\cite{kong2021privacy}.
\section{Experimental Setup}
\begin{figure*}
  \centering
  \subfigure[]{
  \includegraphics[scale=0.06]{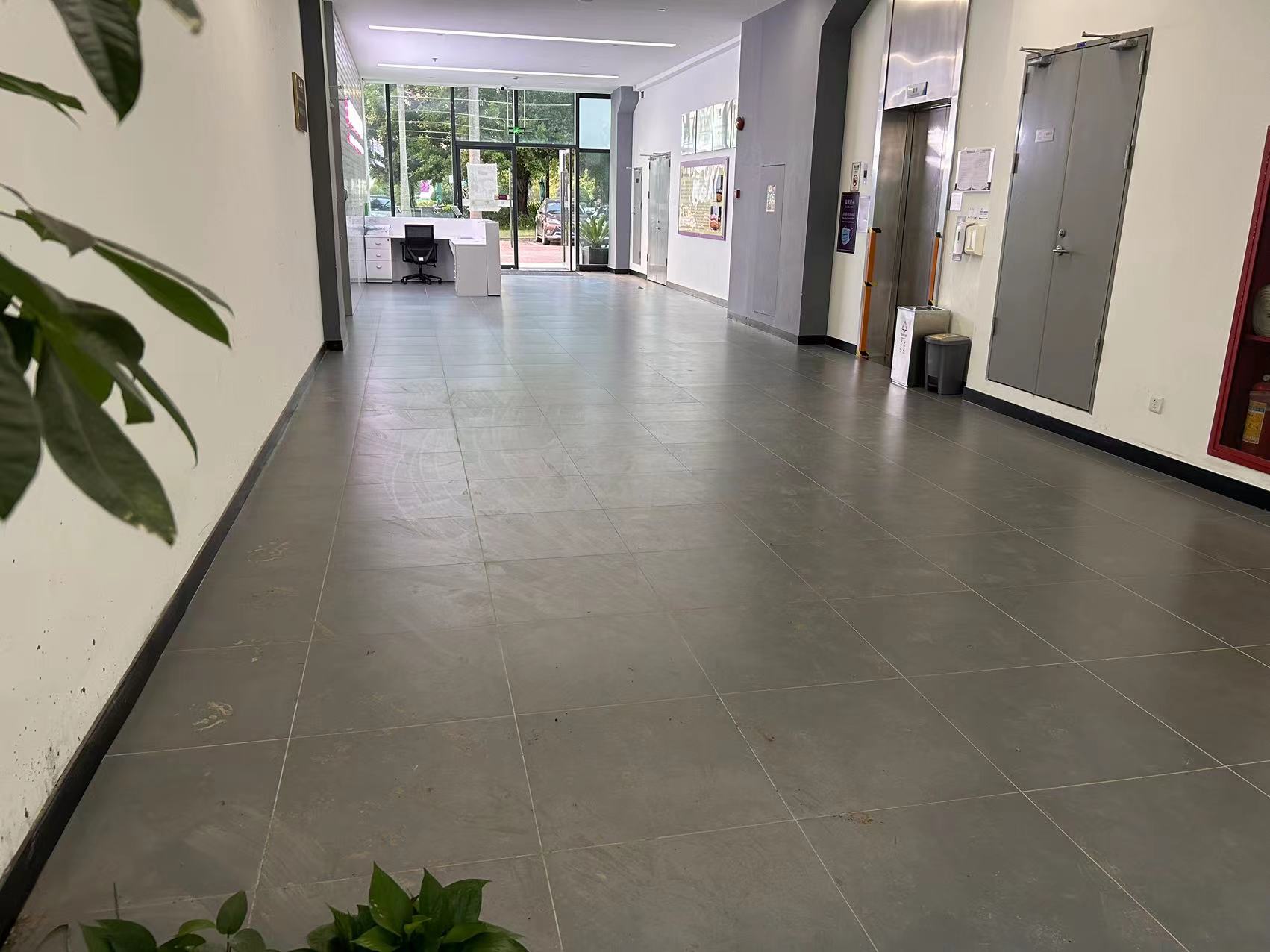}}
  \subfigure[]{
  \includegraphics[scale=0.19]{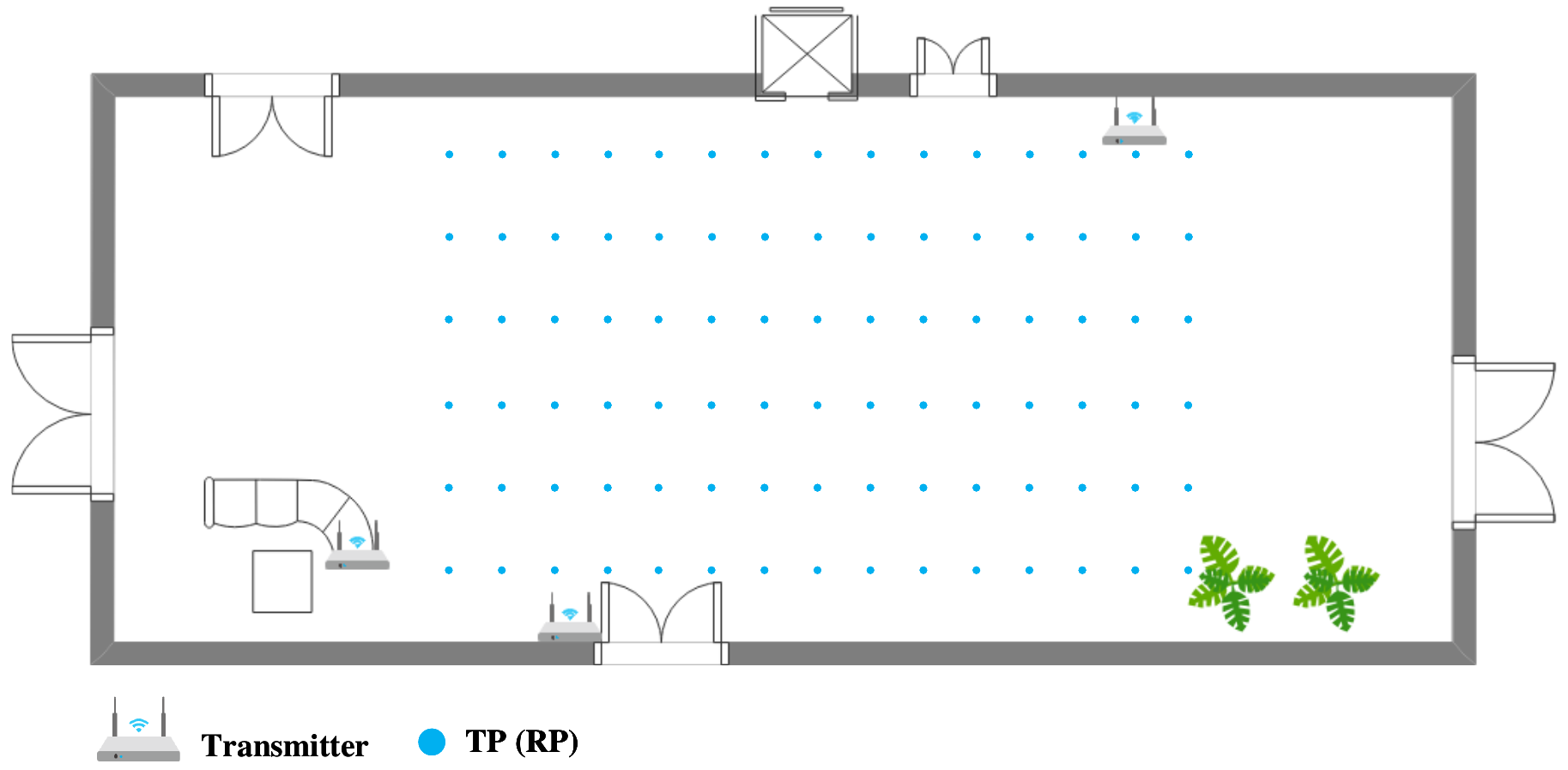}}
  \subfigure[]{
  \includegraphics[scale=0.06]{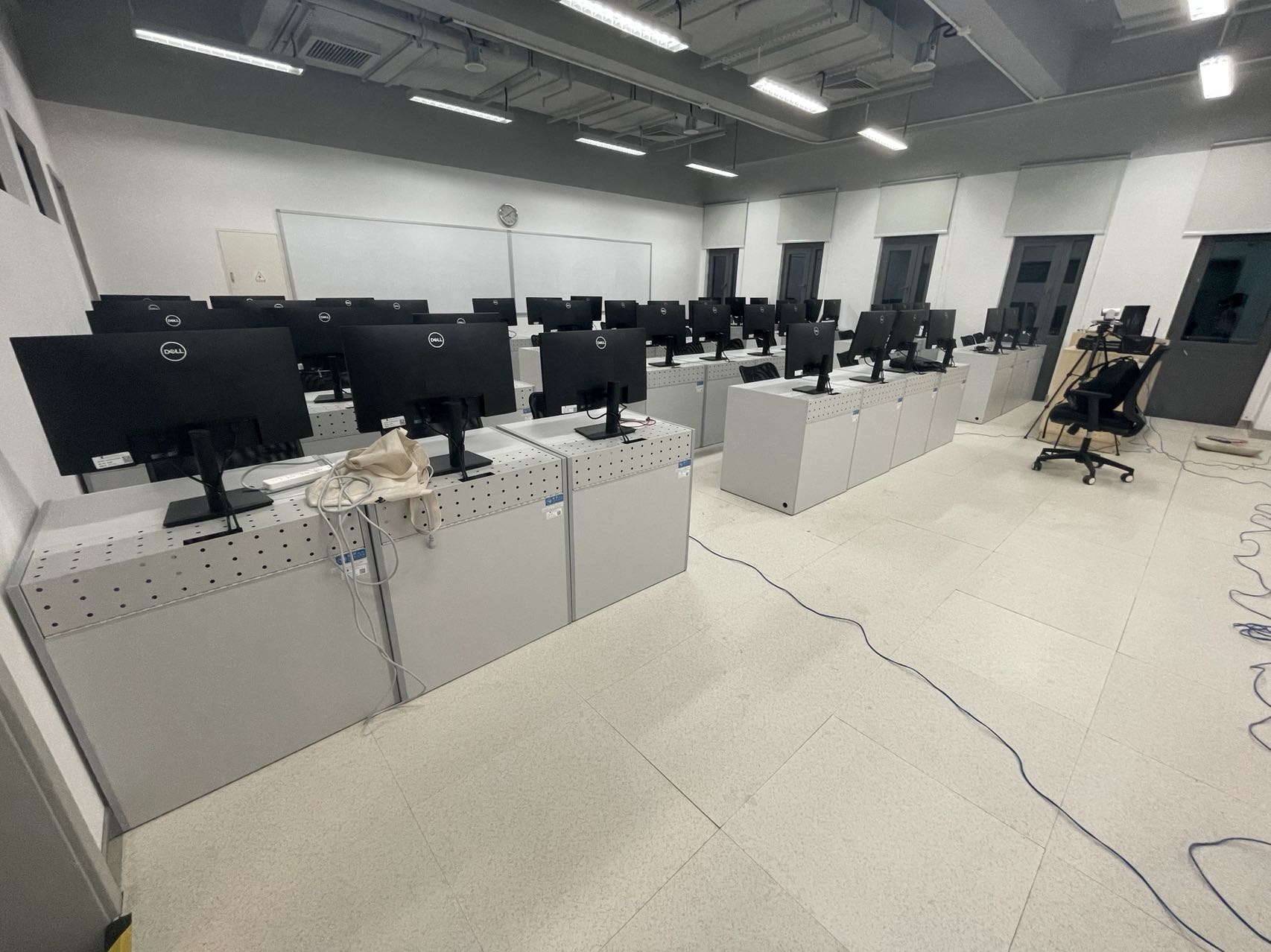}}
  \subfigure[]{
  \includegraphics[scale=0.2]{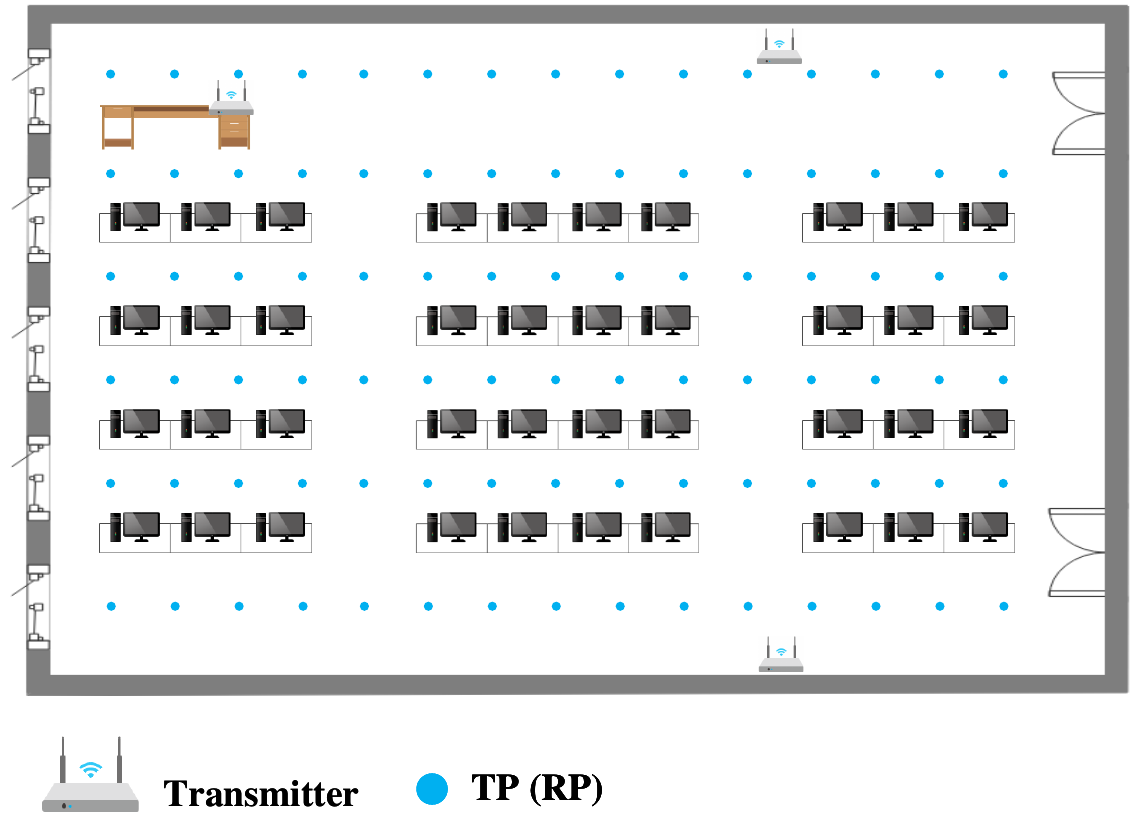}
  }
  
\caption{Photographs and layouts of the two scenarios. (a) The photograph of the hall. (b) The layout of the hall. (c) The photograph of the lab. (d) The layout of the lab. In the historical environments, 90 points are considered RPs, while in the new environment, the 90 points are considered TPs.}
\label{fig:csi}
\end{figure*}
This section describes the data and the baseline models/methods introduced for comparison. Our datasets include both synthetic ones generated from computer simulations and the real ones collected in the site surveys on campus at The Chinese University of Hong Kong, Shenzhen.
\subsection{Synthetic Data from Computer Simulations}
We generate synthetic data from two different perspectives: the 3rd Generation Partnership Project (3GPP) specifications~\cite{3gpp2017study} and the Wireless Insite (WI) platform~\cite{2012Wireless}. The 3GPP specifies propagation models that have been validated by extensive measurement campaigns for typical indoor scenarios. To better reflect real-world propagation conditions, we use the WI platform, which is a suite of ray-tracing models developed by REMCOM that can simulate and predict complex scenarios using advanced electromagnetic processing methods.

From the 3GPP perspective, we classify communication links into line-of-sight (LOS) and non-line-of-sight (NLOS), with probabilities given by Eq.~(1) of \cite{haneda2016indoor}. We present some toy examples by considering multiple path-loss models that capture signal propagation properties in various indoor environments, including:

\begin{itemize} \item Model (a): Vanilla log-distance model; see Eq.~(1) of \cite{bose2007practical}; \item{Model (b): Shopping malls with NLOS dual slopes}; see Eq.~(7) of \cite{haneda2016indoor}; \item{Model (c): Office with mixed LOS and NLOS single slope}~\cite{3gpp2017study};  \item{Model (d): Office with a frequency-dependent path-loss exponent}; see Eq.~(2) and Eq.~(5) of \cite{haneda2016indoor}; \item{Model (e): Shopping malls with mixed LOS and NLOS dual slope}; see Eq.~(2) and Eq.~(8) in~\cite{haneda2016indoor}.
\end{itemize}

We randomly deploy 24 APs operating at 2440 MHz as transmitters. The number of RPs ranges from 10 to 54, and each TP is characterized by its 5-nearest RPs in the signal space. We assume that the data generated from the same path-loss models forms one task.

From the WI perspective, we conduct experiments on six different scenarios generated from the WI platform, in which the layout of the scenarios is designed quite differently. Twenty APs are deployed randomly as transmitters in each scenario. The RSS data generated from scenarios 1, 2, 5 and 6 (S1, S2, S5, S6) are chosen as the training tasks, and the RSS data generated from scenarios 3 (S3) and 4 (S4) are chosen as the test tasks. 

\subsection{Real Data from Site Surveys}
We conduct site surveys to prototype MetaLoc using a Nexus-5 smartphone as the receiver and three different types of routers as transmitters: the ASUS RT-AC86U, TPlink TL-WR885N, and TPlink TL-WR886N. The entire system operates at 5 GHz with a bandwidth of 20 MHz to ensure high-quality wireless channels.

We perform experiments in two different scenarios depicted in Fig.~\ref{fig:csi}. Specifically, Fig.~\ref{fig:csi} (a) and (b) present a hall with a test area of 12 m by 5 m, which is almost empty and can be regarded as a pure LOS environment. Inevitably, there will be people walking and elevators opening or closing. We uniformly selected 90 grid points for data collection, where the distance is 0.6 m between any two adjacent points. We performed a CSI measurement campaign from June 9th to July 14th, 2022. Figure~\ref{fig:csi} (c) and (d) show a lab room of size 10 m by 8 m, where the test area can be regarded as an NLOS environment, since many obstacles, such as desks and computers, are deployed in the room and they blocked the LOS transmission. We selected 90 grid points for data collection. Unlike the hall, the resolution of the grids of lab is inconsistent. Specifically, the interval between two grid points is 0.6 m or 1.2 m. We collected CSI data on five days from July 21st to September 21st, 2022. Each collection is subject to environmental change.

We collected a total of $(k_{spt}+k_{qry}+1)$ CSI images at each grid point in the new environment during the meta-test stage. We collect only one CSI image for the estimated TP and adopt the histogram intersection metric in Eq.~(\ref{eq:HI}) to select its $K$-nearest RPs. We then form the test task $T$, which comprises $K$ RPs with a support set containing $k_{spt}$ CSI images and a query set containing $k_{qry}$ CSI images.
\subsection{Baseline Models/Methods}
We utilize two different neural network architectures, namely multi-layer perceptron (MLP) and CNN, to implement the proposed MetaLoc framework. Our choice of these specific neural networks stems from their widespread use in the state-of-the-art works. For example, ConFi~\cite{chen2017confi} and CiFi~\cite{cifi} utilized CNNs, while  a simple MLP was adopted in~\cite{ghozali2019indoor}. We compare MetaLoc with the state-of-the-art models/methods as follows:
\begin{itemize}
\item KNN~\cite{bahl2000radar}: We adopt the Euclidean distance metric to select closest $K=5$ RPs for the estimated TP in the signal space. The averaged locations of the selected RPs are then treated as the estimation result.
\item {TL~\cite{weiss2016survey}}: Instead of following the standard procedures in the vanilla MAML during the meta-training stage, we adopt the traditional SGD method to train the neural network and obtain the best network parameters for initializing the new environments.
\item {RI}: We randomly generate a set of network parameters for initializing the new environment. The task format and the hyperparameters of the neural network remain the same as those used in the MetaLoc.
\item {ILCL~\cite{zhu2022intelligent}}: We set the number of incremental steps of the BLS classification regression to 10, while the other hyperparameters remain the same as in ~\cite{zhu2022intelligent}.
\item {ConFi~\cite{chen2017confi}}: We keep the neural network architecture and dataset the same as MetaLoc. Data from historical environments serve as the training set, while data collected in new environment serve as the test set.
\end{itemize}

\section{Experimental Evaluations}
\begin{table*} 
\caption{Localization results in the new environment using real data collected from the site surveys}
\centering
\begin{tabular}{ccccccccc} 
\toprule[1.5pt] 
\multicolumn{1}{c}{\multirow{2}*{\textbf{Methods}}}& \multicolumn{4}{c}{\textbf{Hall}}&\multicolumn{4}{c}{\textbf{Lab}}\\
\multicolumn{1}{c}{}&\textbf{Mean errors (m)}&\textbf{Std (m)}&\textbf{Data}&\textbf{Training Steps}&\textbf{Mean errors (m)}&\textbf{Std (m)}&\textbf{Data}&\textbf{Training Steps}\\  
\hline 
\multicolumn{1}{c}{\textbf{MAML (ours)}}                     & \multicolumn{1}{c}{\textbf{2.11}}          & \multicolumn{1}{c}{\textbf{1.17}}  & \multicolumn{1}{c}{\textbf{3$\times$5}}          & \multicolumn{1}{c}{\textbf{7500}}& \multicolumn{1}{c}{\textbf{3.10}}         & \multicolumn{1}{c}{\textbf{1.43}} & \multicolumn{1}{c}{\textbf{3$\times$5}}          & \multicolumn{1}{c}{\textbf{7500}} \\
\multicolumn{1}{c}{\textbf{MAML-DG (ours)}}                     & \multicolumn{1}{c}{\textbf{2.07}}          & \multicolumn{1}{c}{\textbf{1.11}}  & \multicolumn{1}{c}{\textbf{3$\times$5}}          & \multicolumn{1}{c}{\textbf{2500}}& \multicolumn{1}{c}{\textbf{3.04}}         & \multicolumn{1}{c}{\textbf{1.39}}& \multicolumn{1}{c}{\textbf{3$\times$5}}          & \multicolumn{1}{c}{\textbf{2500}}\\
\multicolumn{1}{c}{\textbf{MAML-TS (ours)}}                     & \multicolumn{1}{c}{\textbf{2.09}}          & \multicolumn{1}{c}{\textbf{1.18}}  & \multicolumn{1}{c}{\textbf{3$\times$5}}          & \multicolumn{1}{c}{\textbf{5000}}& \multicolumn{1}{c}{\textbf{3.09}}         & \multicolumn{1}{c}{\textbf{1.35}}& \multicolumn{1}{c}{\textbf{3$\times$5}}          & \multicolumn{1}{c}{\textbf{5000}}\\
\multicolumn{1}{c}{TL}                     & \multicolumn{1}{c}{2.27}          & \multicolumn{1}{c}{1.27}  & \multicolumn{1}{c}{3$\times$5}          & \multicolumn{1}{c}{$7500$}& \multicolumn{1}{c}{3.97}         & \multicolumn{1}{c}{1.99}&\multicolumn{1}{c}{3$\times$5}          & \multicolumn{1}{c}{$7500$}\\
\multicolumn{1}{c}{RI}                     & \multicolumn{1}{c}{2.59}          & \multicolumn{1}{c}{1.29}  &\multicolumn{1}{c}{3$\times$5}          & \multicolumn{1}{c}{$7500$}&\multicolumn{1}{c}{4.19}         & \multicolumn{1}{c}{1.95}&\multicolumn{1}{c}{3$\times$5}          & \multicolumn{1}{c}{$7500$}\\
\multicolumn{1}{c}{ConFi}                     & \multicolumn{1}{c}{2.89}          & \multicolumn{1}{c}{0.48}  &\multicolumn{1}{c}{260$\times$90}          & \multicolumn{1}{c}{$7500$}& \multicolumn{1}{c}{3.53}         & \multicolumn{1}{c}{0.47}&\multicolumn{1}{c}{260$\times$90}          & \multicolumn{1}{c}{$7500$}\\
\multicolumn{1}{c}{ILCL}                     & \multicolumn{1}{c}{3.61}          & \multicolumn{1}{c}{2.06}  &\multicolumn{1}{c}{$3\times90$}          & \multicolumn{1}{c}{$7500$}& \multicolumn{1}{c}{3.48}         & \multicolumn{1}{c}{1.62}&\multicolumn{1}{c}{$3\times90$}          & \multicolumn{1}{c}{$7500$}\\
\multicolumn{1}{c}{KNN}                     & \multicolumn{1}{c}{2.73}          & \multicolumn{1}{c}{1.35}  &\multicolumn{1}{c}{$10\times90$}     & \multicolumn{1}{c}{$/$}&\multicolumn{1}{c}{3.35}         & \multicolumn{1}{c}{1.42}&\multicolumn{1}{c}{$10\times90$}          & \multicolumn{1}{c}{$/$}\\
\bottomrule[1.5pt]
\end{tabular}
\label{tab:results}
\end{table*}

In this section, we formulate localization as a regression problem and present some preliminary results of the toy examples using synthetic data generated from computer simulations. We also formulate localization as a classification problem and verify the efficacy of the proposed MetaLoc based on real site-surveyed data collected from hall and lab shown in Fig.~\ref{fig:csi}.

\subsection{Toy Examples Based on Computer Simulations}
 For each task in the simulation, we assume the number of support samples $k_{spt}$ is approximately 100, and the number of query samples $k_{qry}$ is 30.  As for the training procedure, we set the step size of the inner loop to be $\alpha = 0.0001$ and the step size of the outer loop $\beta = 0.001$. A neural network architecture of MLP, consisting of four hidden layers, is considered to formulate localization as a regression problem, with observed RSS fingerprints shown in Fig.~\ref{fig:input_data} as input and the corresponding locations as output.
\subsubsection{Convergence Speed} 
\begin{figure}[tbh]
\begin{centering}
\includegraphics[scale=0.45]{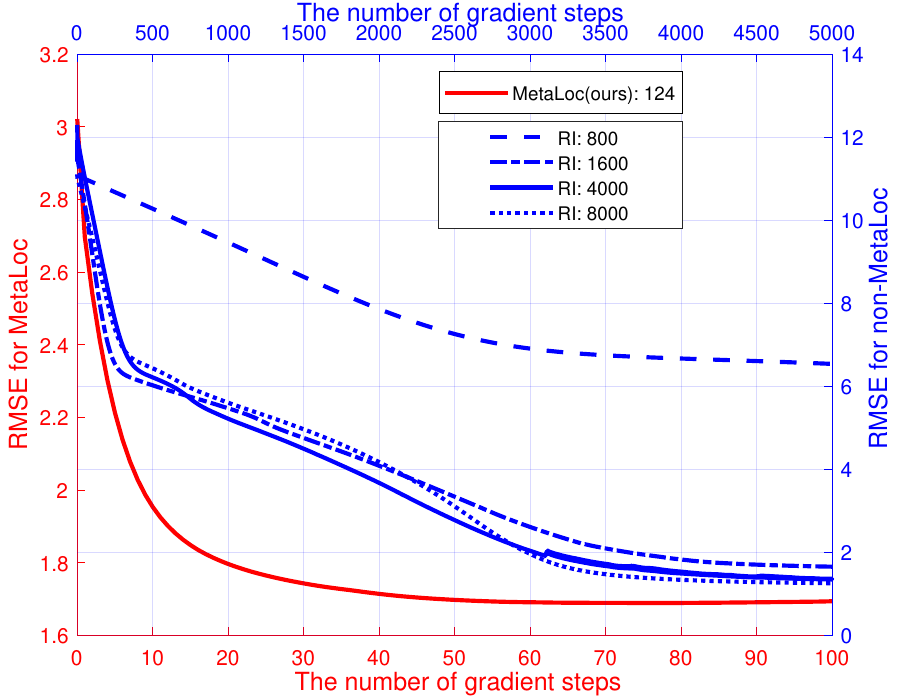}
\par\end{centering}
\caption{RMSE convergence comparisons between the MetaLoc and the baseline method RI, which is trained using the same neural network architecture but with the random initialization of the network parameters. In the double-axis system, the red-axis system represents MetaLoc with 124 training data, while the blue-axis system represents RI with 8000, 4000, 1600 and 800 training data.}
\label{fig:Double-axis-system:}
\end{figure}

To test the convergence speed of MetaLoc, we exploit a vanilla log-distance path-loss model operating at different std values to generate synthetic data with the experimental settings given in Table I of \cite{gao2022metaloc}, which lists four training tasks and one test task with various transmit powers, path-loss exponents, antenna gains and noise levels. In the double-axis system as shown in  Fig.~\ref{fig:Double-axis-system:}, the red-axis system represents MetaLoc, while the blue-axis system represents the baseline method RI that was trained using the same neural network architecture but with random initialization of the network parameters. Figure~\ref{fig:Double-axis-system:} presents the relationship between RMSE and the number of gradient steps with different training data size. Specifically, the red curve shows that MetaLoc converges much faster than those RI curves. Moreover, Fig.~\ref{fig:Double-axis-system:} shows that the RI method can be largely affected by the training data size, i.e., the RI performance improves with an increasing scale of training data. When the data size rises to 8000 samples, the localization performance becomes saturated, reaching a level slightly inferior to that of MetaLoc but the latter merely requires only 124 data samples for training. Compared with the traditional fingerprinting methods that strongly rely on the large amount of data collected in the target environment, the MetaLoc framework exploits the existing database built for a batch of different scenarios.

Upper bound of $ER\left(\bm{\theta}_{T}{(Q)}\right)$  in Theorem I explains the rapid adaptation of  MetaLoc from the perspective of parameter space:
the smaller distance
between $\bm{\theta}^{*}$ and $\bm{\theta}_{T}^{*}$
, the smaller excess risk, guaranteeing good test performance of
$\bm{\theta}_{T}(Q)$ on its corresponding task $T$. We further verify MetaLoc by computing Euclidean distance  between $\bm{\theta}^{*}$ and $\bm{\theta}_{T}^{*}$ and the resultant distance is 0.48, while for RI cases, the average distance between the randomly initialized parameters and $\bm{\theta}_{T}^{*}$ is 2.12, which show meta-parameters $\bm{\theta}^{*}$ locate close to the optimal parameters $\bm\theta^{*}_{T}$ and thus facilitate rapid adaptation to new scenarios. 

\begin{figure}[t]
\begin{centering}
\includegraphics[scale=0.65]{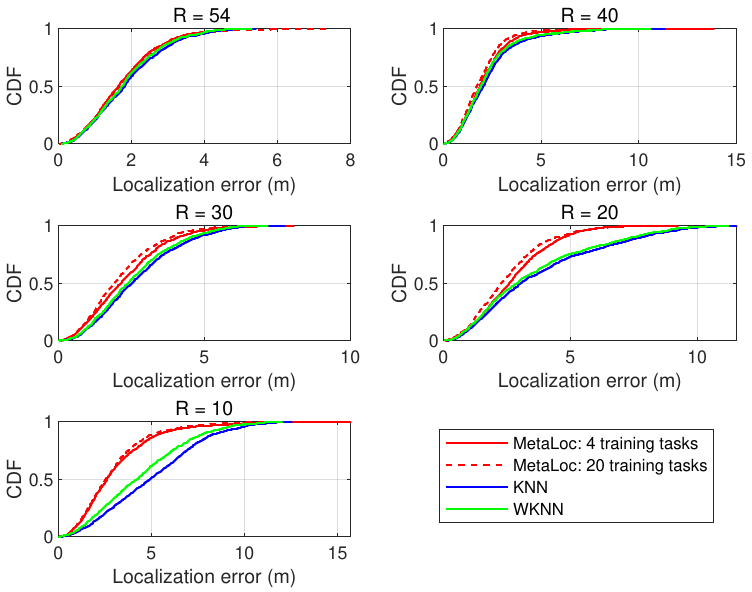}
\caption{CDFs of localization errors (m)
under different RP numbers ($R$).}\label{fig:CDF}
\par\end{centering} 
\end{figure}

\subsubsection{Localization Accuracy} 
We randomly sample 1000 test samples in the test task and quantify the localization errors in terms of the cumulative distribution function (CDF). Figure~\ref{fig:CDF}  presents the CDF of the localization errors versus different RP numbers ($R$), where $R = 10, 20, 30, 40, 54$. We observe that more RPs deployed in the scenario can promote localization accuracy due to the abundant characteristics of the multipath channel fed into the network. More specifically, when the number of RP $R = 54$, the localization results of the traditional KNN, WKNN and our MetaLoc are nearly the same. As $R$ decreases, the performance gap becomes more significant, and MetaLoc shows higher resistance to performance degradation. When $R$ decreases to 10, the probability of the localization errors of MetaLoc, WKNN and KNN being less than $5$ m are 0.89, 0.65 and 0.51, respectively. MetaLoc presents better accuracy in the severe case of $R=10$. The above findings indicate that MetaLoc outstandingly reduces the dependence on a large number of RPs and shows the best cost-effectiveness in constructing a fingerprint database.

\subsubsection{Impact of environment-specific meta-parameters}
\begin{figure}[t]
	\begin{centering}
		\includegraphics[scale=0.55]{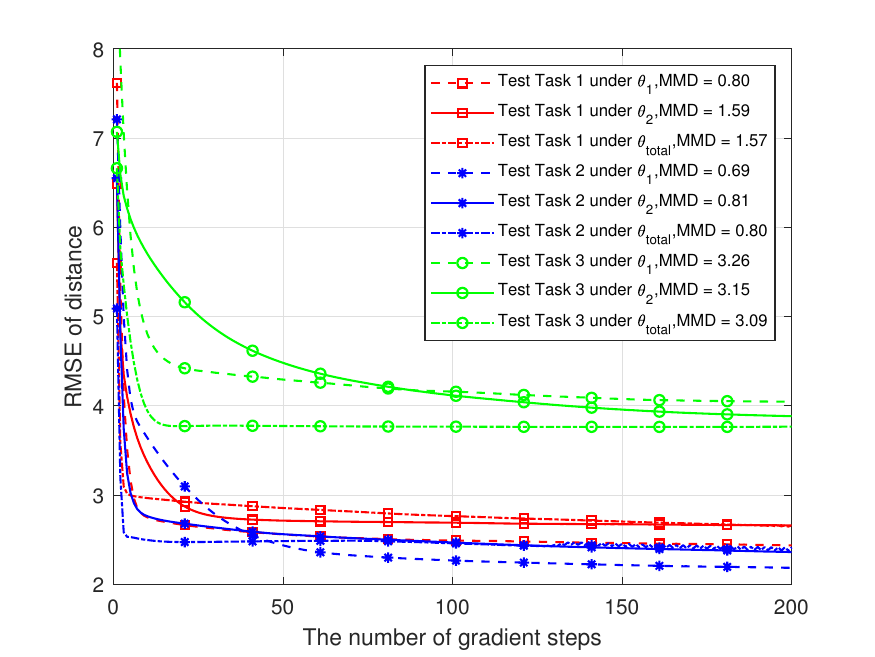}
		\par\end{centering}
	\caption{The test result comparisons among three different meta-parameters, where $\bm{\bm{\theta}}_{1}$, $\bm{\bm{\theta}}_{2}$, and $\bm{\bm{\theta}}_{total}$ represent the well-trained meta-parameters based on two environments, i.e., Environment One, Environment Two, and the total training tasks listed in Table II of reference~\cite{gao2022metaloc}, respectively.\label{fig:MMDall}}
\end{figure}

We utilize multiple path-loss models from 3GPP to characterize different environments instead of the vanilla log-distance model. In Table II of~\cite{gao2022metaloc}, various training tasks with different transmit powers, room sizes, and noise levels are listed. We divide them into three environments based on the path-loss model generation. Environment One is generated from Model (b), Environment Two from Model (c), and Environment Three from Model (d). For the test tasks, we simulate scenarios with a square layout of size 10 m by 10 m and a transmit power of $P_{t}=10$ dBm. Test Task One is simulated from Model (e) with $\sigma_{LOS} = 3$, $\sigma_{NLOS} = 6.26$; Test Tasks Two and Three are generated from Model (b) with $\sigma_{NLOS} = 5$ and $\sigma_{NLOS} = 15$, respectively. It should be noted that both Model (b) and Model (e) simulate shopping mall scenarios.
MMD measures the average difference between each test task and the training environments, as shown in Fig.~\ref{fig:MMDall}. 

In Fig.~\ref{fig:MMDall}, $\bm{\theta}_1$ and $\bm{\theta}_2$ represent the environment-specific meta-parameters trained on Environment One and Environment Two, and $\bm{\theta}_{total}$ are trained on all three environments listed in Table II of~\cite{gao2022metaloc}. It can be observed that MMD reflects the quality of the test performance, with test tasks with smaller MMD values achieving better localization results. The learned meta-parameters exhibit rapid convergence on multiple test tasks, indicating good generalization ability to new scenarios. Furthermore, in the test tasks with a common noise standard deviation proposed in 3GPP, the environment-specific meta-parameters $\bm{\theta}_{1}$ outperform $\bm{\theta}_{2}$ and $\bm{\theta}_{total}$. This is because Environment Two and Environment Three, both generated from office scenarios, cannot provide much specific assistance for test tasks simulated in malls and may even introduce outliers. However, as the noise standard deviation added to the test tasks becomes significantly large, there is no noticeable improvement in the test tasks on $\bm{\theta}_{1}$, which has limitations in special cases with extreme noise standard deviation inputs. Overall, these results indicate that MMD can provide preliminary information about the task similarity to assist in selecting environment-specific meta-parameters and to facilitate further improvement in localization accuracy.

\subsubsection{Wireless Insite (WI)}
\begin{figure}[tbh]
\begin{centering}
\includegraphics[scale=0.55]{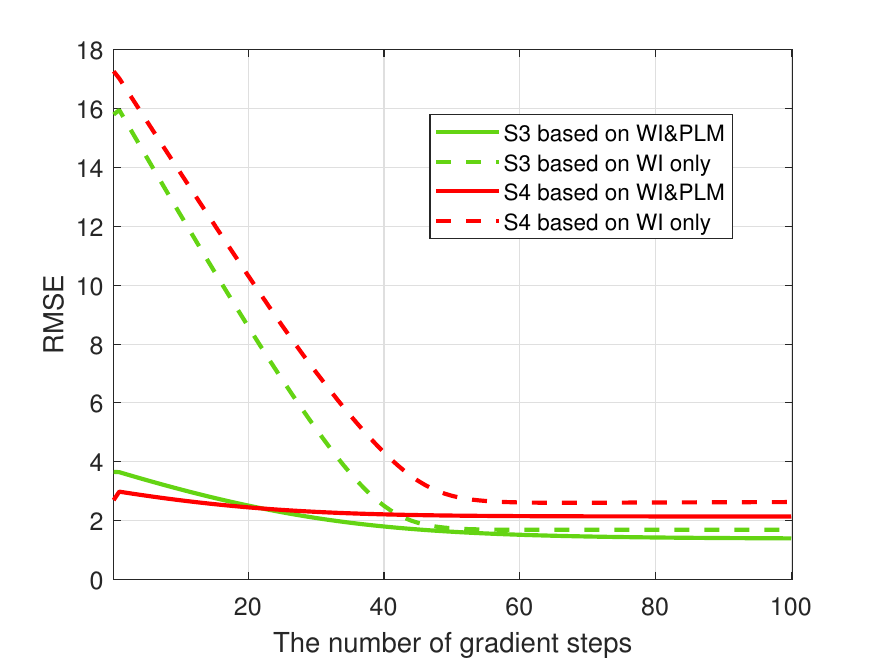}
\par\end{centering}
\caption{Comparison results for data generated from the WI platform where PLM represents path-loss models.\label{fig:Comparison-results-for}}
\end{figure}

Figure~\ref{fig:Comparison-results-for} compares the RMSE results for the data
generated from the WI platform under the well-trained parameters obtained by the proposed framework. We consider four cases as shown in Fig.~\ref{fig:Comparison-results-for}. Specifically, S3, S4 based on WI\&PLM represent the well-trained initialization from the path-loss models (abbreviated as PLM in the figure) and WI platform, while S3, S4 based on WI represent the well-trained initialization from the WI platform only. We observe that the results converge extremely fast in the first 50 iterations under all four cases. Interestingly, the solid lines are observed to converge more quickly than the dot lines. A possible explanation for
this might be that the well-trained initialization from the path-loss models  and WI have more knowledge of the channel features than those trained only over
the WI data. Our findings confirm that simulation data can provide extra support in alleviating the     data-hungry nature of data-driven localization methods. 

\subsection{Comprehensive Results Based on Real Site Surveys}
 For each task in the site surveys, we set the number of points $N=10$, the number of support data $k_{spt}=3$ and the number of query data $k_{qry}=5$.  A CNN including five convolution layers, pooling layers and a fully connected layer, is considered a neural network architecture that
formulates localization as a classification problem. The observed CSI fingerprints shown in Fig.~\ref{fig:csi-ff} are taken as input and the location probabilities as the
output. During the meta-training stage, the step size of the inner loop $\alpha$ and the outer loop $\beta$ are set to 0.01 and 0.001, respectively. In addition, we set the number of gradient descent steps for the inner loop to 5. We collect data on five days, four of which are used for training tasks (i.e., historical environments), and data collected on the fifth day are used for test tasks (i.e., new environments).

\subsubsection{Localization errors}
\begin{figure}
\centering
\subfigure[Hall]{\includegraphics[scale=0.285]{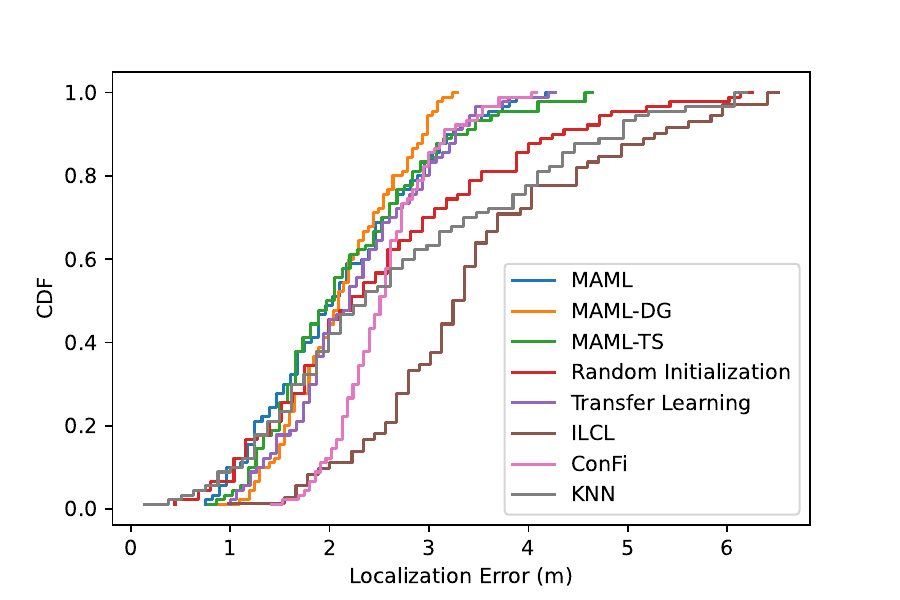}}
\subfigure[Lab]{\includegraphics[scale=0.285]{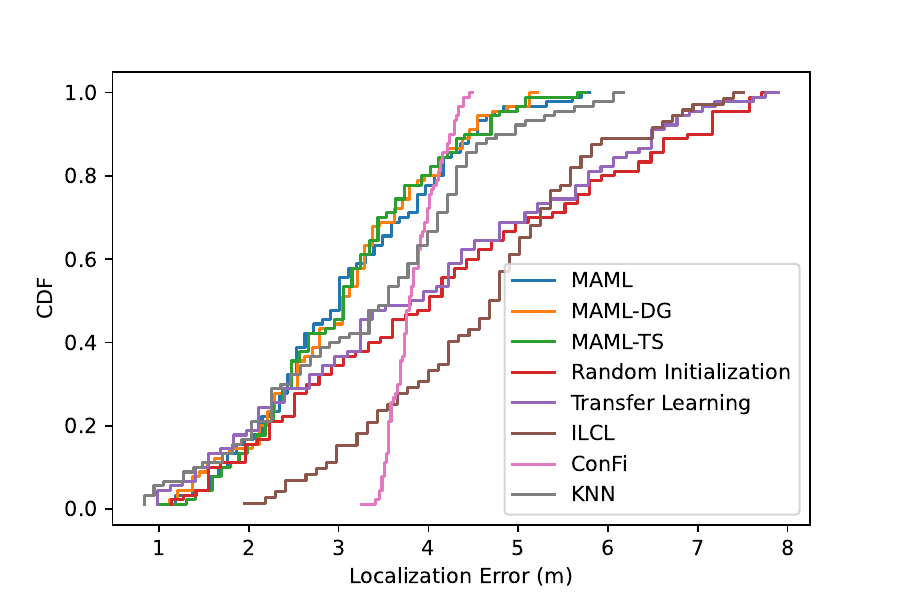}}
\caption{CDF comparison of average localization errors over different dates in the hall and the lab.}
\label{fig:CDF-nlos}
\end{figure}

Table~\ref{tab:results} and Fig.~\ref{fig:CDF-nlos} illustrate the localization results of the different methods using real data in the site surveys. We observe that the proposed MAML-DG shows the lowest mean errors of 2.07 m and 3.04 m in the hall and lab, respectively. There are two reasons for the decrease in the localization performance in the lab. First, the propagation environment of the lab is complex with office facilities, and the obstacles exacerbate the multiple path effects. Second, the grid resolution in the lab is inconsistent with that in the hall due to the setup of office cubicles, with a spacing of 1.2 m in the former and 0.6 m in the latter. The coarse resolution comes with a performance penalty in the lab. Performance degradation is particularly evident in the baseline method RI, because the initialization was not effectively learned in  harsher environments during the training stage. Moreover, in less complex environment such as the hall scene, the performance gap between TL and our MetaLoc is minor. But in a more complex lab environment, the difference becomes noticeable. This is because environmental complexity magnifies the conflicts among the learned knowledge, impairing the performance of TL. In contrast, the inherent learning-to-learn capability of meta-learning effectively mitigates these knowledge conflicts, thereby maintaining performance even in more complex scenes.

Besides, it is noticed that the baseline method ILCL shows improvement in the lab, but the positioning error are somewhat high. The ILCL method is susceptible to the regularization term and data volume. Even when the regularization term is adjusted to the optimal value, the model tends to overfit when the number of CSI images of each test point is too small. Compared to ILCL, MetaLoc is more robust and can quickly adapt to a new environment with only a small sample size. The competing method ConFi presents the smallest variance of the localization errors, but requires a large amount of data for training in the new environment. Meanwhile, the lowest localization error of ConFi is up to 1.8 m, which is far higher than any other method. The performance of KNN lies in the middle, which is as expected since KNN has poor tolerance to outliers generated due to the fast-changing environments. 

In the following, we will delve deeper into the convergence of the proposed framework during both the meta-training and meta-test stages, illustrated in Fig.~\ref{fig:loc-steps-nlos} and Fig.~\ref{fig:test-error}, respectively. The former demonstrates the impact of the meta-parameters during the meta-training stage, while the latter showcases the efficacy of the well-trained meta-parameters in adapting to a new environment.

\subsubsection{Convergence of localization errors}
\begin{figure}
\centering
\subfigure[Hall: June 23rd, 2022]{\includegraphics[scale=0.3]{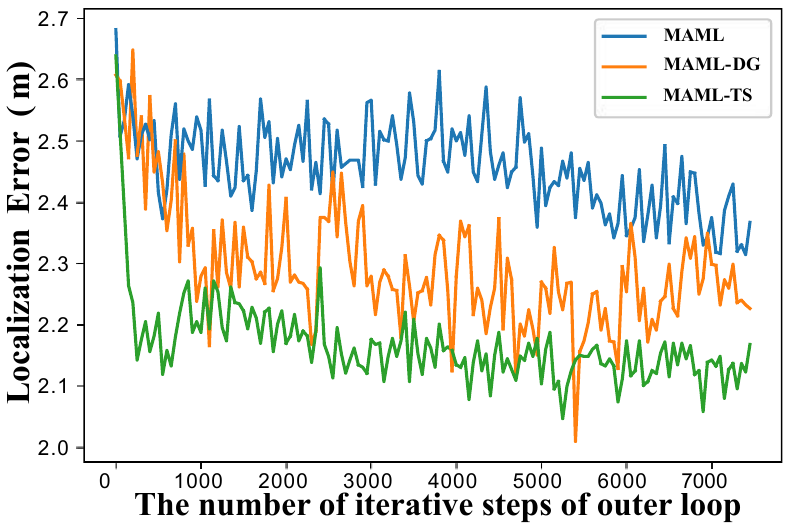}}
\subfigure[Hall: June 30th, 2022]{\includegraphics[scale=0.3]{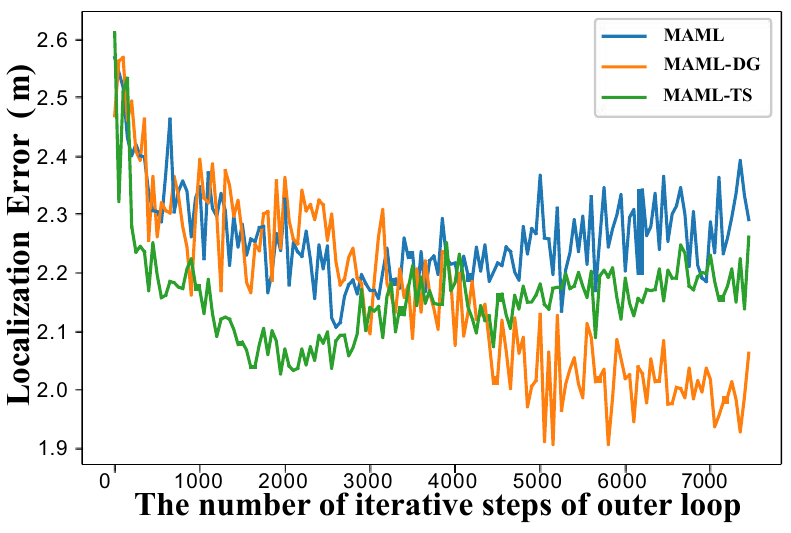}}\\
\subfigure[Lab: July 21st, 2022]{\includegraphics[scale=0.3]{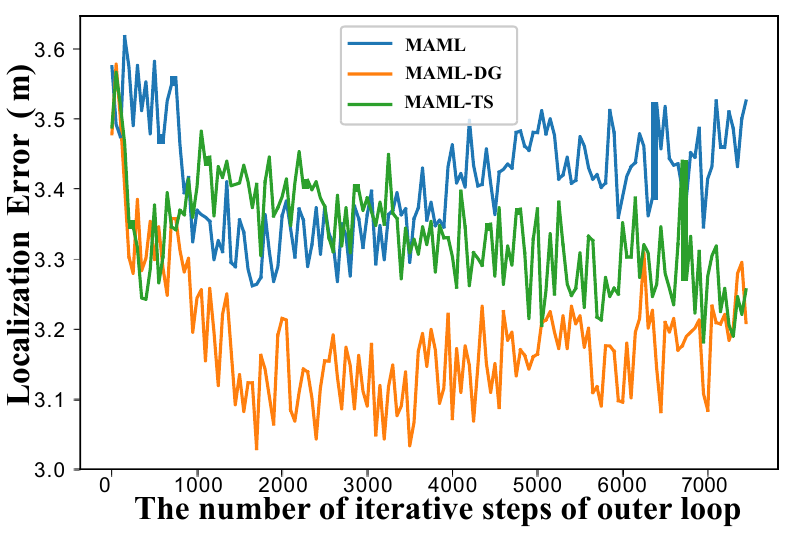}}
\subfigure[Lab: September 25th, 2022]{\includegraphics[scale=0.3]{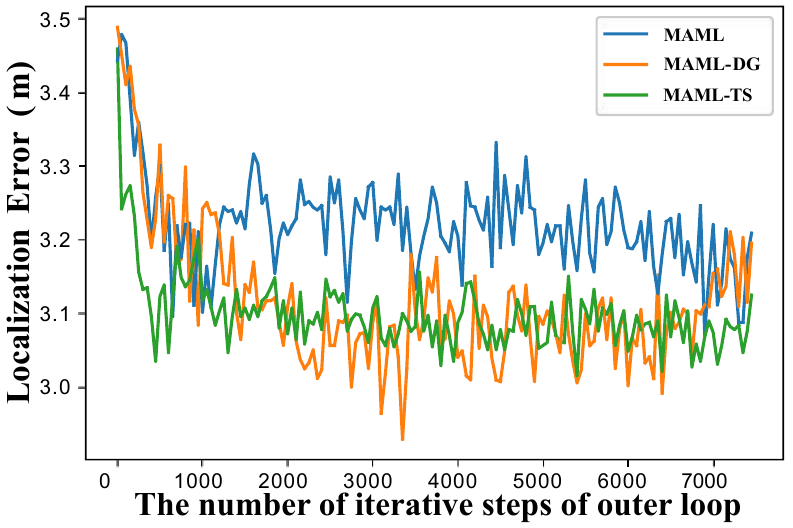}}
\caption{Convergence comparison of the localization errors in the hall and the lab.}
\label{fig:loc-steps-nlos}
\end{figure}

During the meta-training stage, the localization errors with respect to the number of iterative steps of the outer loop for the three paradigms of MetaLoc, i.e., vanilla MAML, MAML-DG and MAML-TS, are shown in Fig.~\ref{fig:loc-steps-nlos}. Overall, MAML-TS and MAML-DG can achieve faster convergences and smaller localization errors than vanilla MAML. However, MAML-TS sometimes suffers from overfitting since it is only trained in one specific environment. In contrast, as shown in Fig.~\ref{fig:loc-steps-nlos} (b), MAML-DG effectively avoids the overfitting problem because it captures the domain differences from the various environments.
\subsubsection{Convergence of test errors}
\begin{figure}
\centering
\subfigure[Hall]{\includegraphics[scale=0.3]{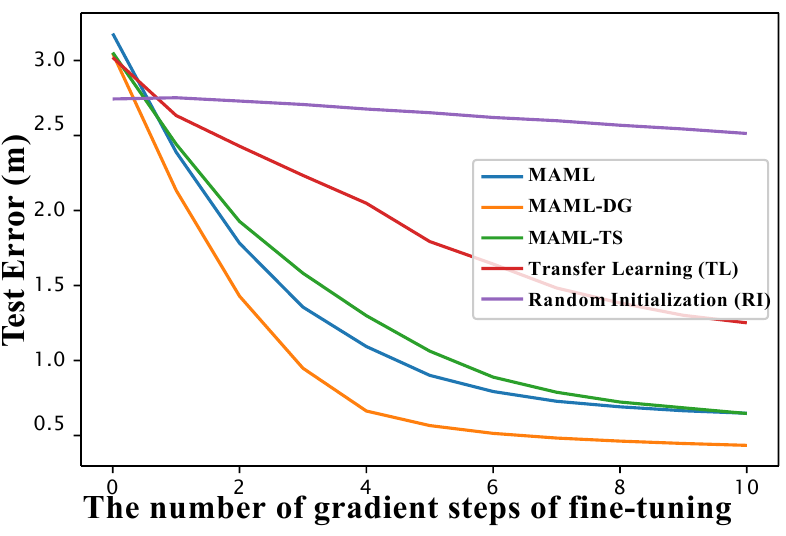}}
\subfigure[Lab]{\includegraphics[scale=0.3]{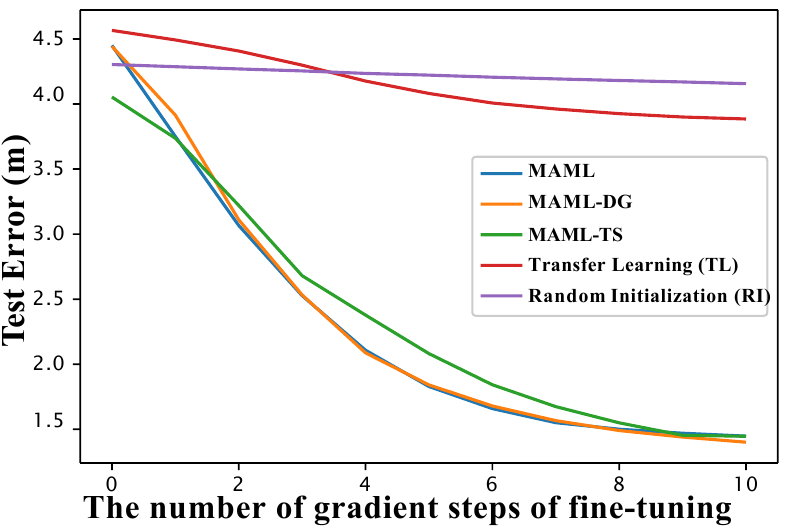}}
\caption{Convergence comparison of test errors in the hall and the lab.}\label{fig:test-error}
\end{figure}

During the meta-test stage in a new environment, the well-trained meta-parameters are obtained, and the convergence results of the test errors with respect to the number of gradient steps are shown in Fig.~\ref{fig:test-error}. As demonstrated in the figure, the proposed MetaLoc outperforms the baseline methods TL and RI, requiring only a few gradient steps to converge to satisfactory performance. This showcases the extraordinary environmental adaptation abilities of MetaLoc. Moreover, our proposed MAML-DG requires fewer gradient steps to converge than both MAML and MAML-TS.

\subsection{Further Discussions about MetaLoc and TL}
MetaLoc and TL are both techniques that apply previous learning experiences to new tasks. TL primarily imparts existing knowledge, while MetaLoc conveys learning strategies across diverse tasks. TL can encounter the problem of negative transfer when there is conflict in the shared knowledge. In dynamic environments for wireless localization, where interference, noise, and physical obstructions continuously change the channel conditions, the learning-to-learn ability of MetaLoc, becomes particularly useful. On the contrary, TL strongly depends on the knowledge specific to training environments, which may not work effectively to new scenarios, especially when the condition is very complex, as shown in Fig.~\ref{fig:CDF-nlos} (b).

\section{Future Work and Challenges Towards 6G}
The improvements of MetaLoc in accuracy, robustness, and cost-effectiveness foster the development of further research, particularly in the forthcoming 6G era. Considering large scale and diverse scenarios in 6G, the learning-to-learn capability of MetaLoc
becomes a spotlight in this emerging research frontier, paving the way for intelligent, adaptive networks. In this section, we will shed some light on a few representative scenarios and associated challenges.
\subsection{Future Work}
\textbf{Joint Communication and Sensing (JCAS):} In 6G communication systems, the incorporation of higher frequency bands, wider bandwidths, and massive antenna arrays not only enhances traditional communication services, but also supports high-precision, high-resolution sensing~\cite{9971740}, etc. MetaLoc, utilizing the wireless network as a sensor, is capable of rapidly adapting to new environments for precise user localization. This location information, in turn, can improve communication performance in our future work,  placing MetaLoc as a cost-effective example of JCAS, especially for dynamic environments. In the future, we plan to extend this framework to fuse wireless signal data that carry information about human postures and the location information, making MetaLoc suitable for applications like smart homes and remote surgeries.

\textbf{Digital Twins:} Digital twins offer virtual representations of physical systems through ML, data analytics, and multi-physics simulation~\cite{9711524}. As 6G technology evolves towards self-sustainability and proactive online learning, it aims to minimize human intervention and thus reduce costs. With its inherent capacity of learning-to-learn, MetaLoc is well-suited to this context. In the framework of digital twins, when deviations arise between the simulated and actual environments, model adjustments are necessary. Utilizing accumulated historical data, MetaLoc can swiftly facilitate model recalibration in a cost-effective way. Our research validates that meta-parameters derived from 3GPP-defined environments effectively facilitate learning in WI-defined scenarios, thus reinforcing the capabilities of digital twins. This interplay between MetaLoc and digital twins illuminates the potential of such a combination in advancing 6G self-sustaining wireless systems.

\subsection{Challenges}
The procurement of high-quality and substantial training data in the dynamic landscape of 6G networks presents considerable challenges. Each environment, with its unique attributes and inherent interference, may result in data of different forms. The mixture of this heterogeneous data into a coherent system necessitates the utilization of advanced methodologies such as multimodal data fusion techniques~\cite{8269806}. Moreover, before using the data in these algorithms, thorough preprocessing is typically a necessary procedure. This includes steps such as noise filtering, outlier removal, and data normalization, which collectively ensure the preservation of data integrity and the enhancement of its utility~\cite{CRISLoc}. To distill valuable features from the raw data, principal component analysis (PCA)~\cite{zhou2017device} is often utilized. Beyond the realm of traditional techniques, leveraging advanced feature construction methods driven by deep learning~\cite{wei2021data,zhang2022fast} could provide a more sophisticated level of feature representation.  Moreover, data privacy regulations may potentially restrict the volume and type of data accessible for MetaLoc, and some potential solutions might involve synthetic data generation~\cite{creswell2018generative}, federated learning~\cite{yin2020fedloc}, each solution requiring further exploration when combined with MetaLoc. 

\section{Conclusion}
The MetaLoc framework proposed in this paper provides a robust solution for fingerprinting-based localization in dynamic and uncertain environments, ensuring efficient wireless localization. The framework includes two MetaLoc paradigms that leverage past knowledge to learn meta-parameters, resulting in significant improvements in localization accuracy, convergence, and cost-effectiveness. Additionally, MetaLoc is highly versatile and can be applied to any model that uses gradient-based training, making it compatible with a range of classification and regression problems. With its superior performance, MetaLoc has promising potential for large-scale deployment in challenging 5G/6G scenarios as we mentioned in Section VI. We believe that the philosophy of \textbf{learning-to-learn} will extend far beyond the scope of wireless localization, holding substantial relevance for the future large-scale intelligent systems.

\section*{Acknowledgement}
The work of Shuguang Cui was supported in part by NSFC with Grant No. 62293482, the Basic Research Project No. HZQB-KCZYZ-2021067 of Hetao Shenzhen-HK S\&T Cooperation Zone, the National Key R\&D Program of China with grant No. 2018YFB1800800, the Shenzhen Outstanding Talents Training Fund 202002, the Guangdong Research Projects No. 2017ZT07X152 and No. 2019CX01X104, the Guangdong Provincial Key Laboratory of Future Networks of Intelligence (Grant No. 2022B1212010001), the Shenzhen Key Laboratory of Big Data and Artificial Intelligence (Grant No. ZDSYS201707251409055), and the Key Area R\&D Program of Guangdong Province with grant No. 2018B030338001.

The work of Feng Yin was supported by the NSFC under Grant No. 62271433 and No. 92067202, and in part by Guangdong Zhujiang Project under Grant No. 2017ZT07X152 and by Shenzhen Science and Technology Program under Grant No. JCYJ20220530143806016.

The work of Qinglei Kong was supported in part by the NSFC under Grant No. 62202127 and the
Guangdong Provincial Key Laboratory of Future Networks of
Intelligence (Grant No. 2022B1212010001).

\bibliographystyle{IEEEtran}
\bibliography{reference}

\begin{IEEEbiography}[{\includegraphics[width=1in,height=1.25in,clip,keepaspectratio]{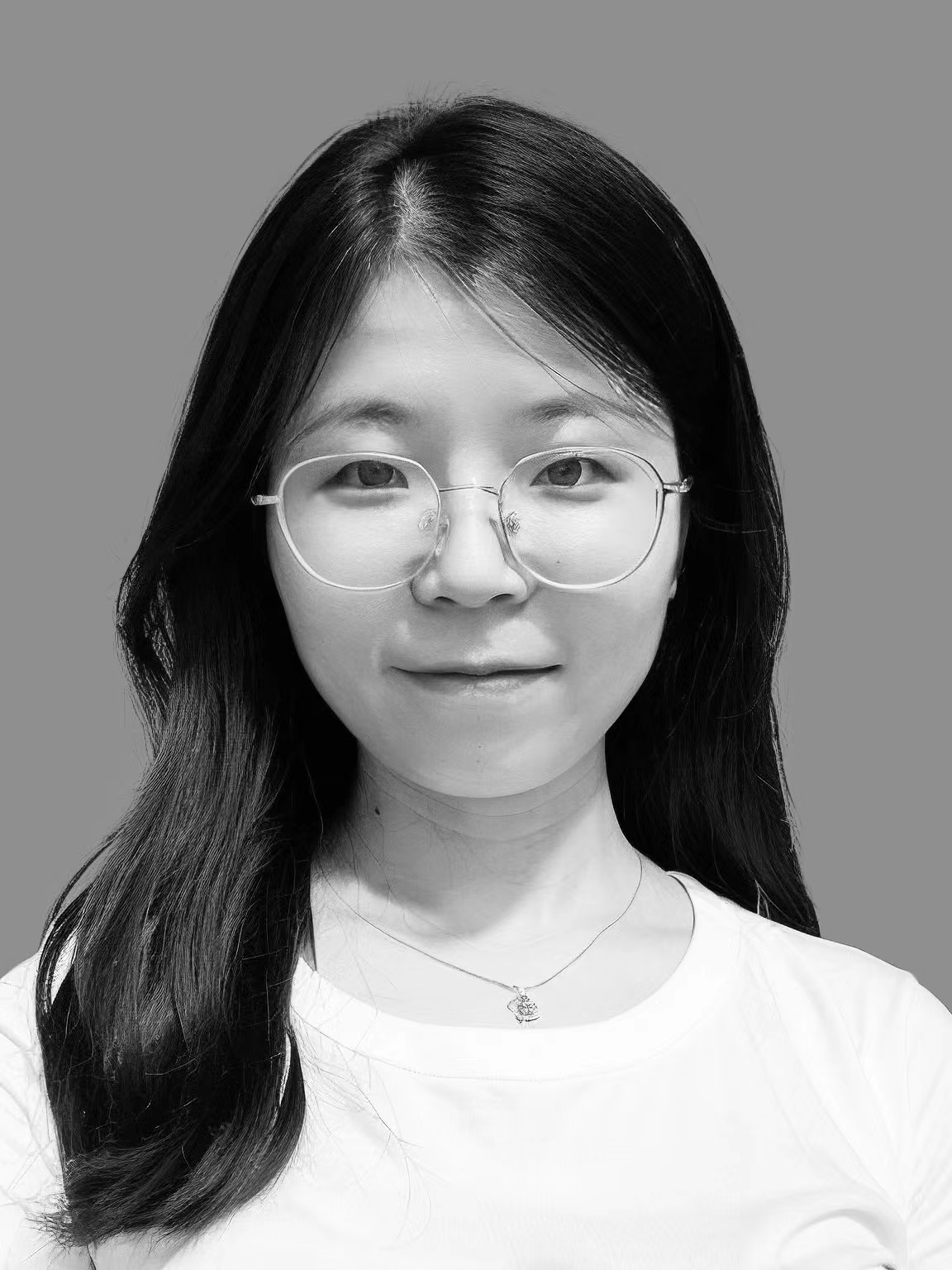}}]{Jun Gao} obtained her B.Eng. degree in communication engineering from Nanchang University, Nanchang, China, in 2016. She received her M.Eng. degree in electronic and communication engineering from Beijing University of Posts and Telecommunications, Beijing, China, in 2019. She is currently pursuing her Ph.D. degree with the Future Network of Intelligence Institute (FNii), and also with the School of Science and Engineering (SSE), The Chinese University of Hong Kong, Shenzhen, China. Her research interests focus on machine learning, as well as wireless sensing and localization.
\end{IEEEbiography}
\begin{IEEEbiography}[{\includegraphics[width=1in,height=1.25in,clip,keepaspectratio]{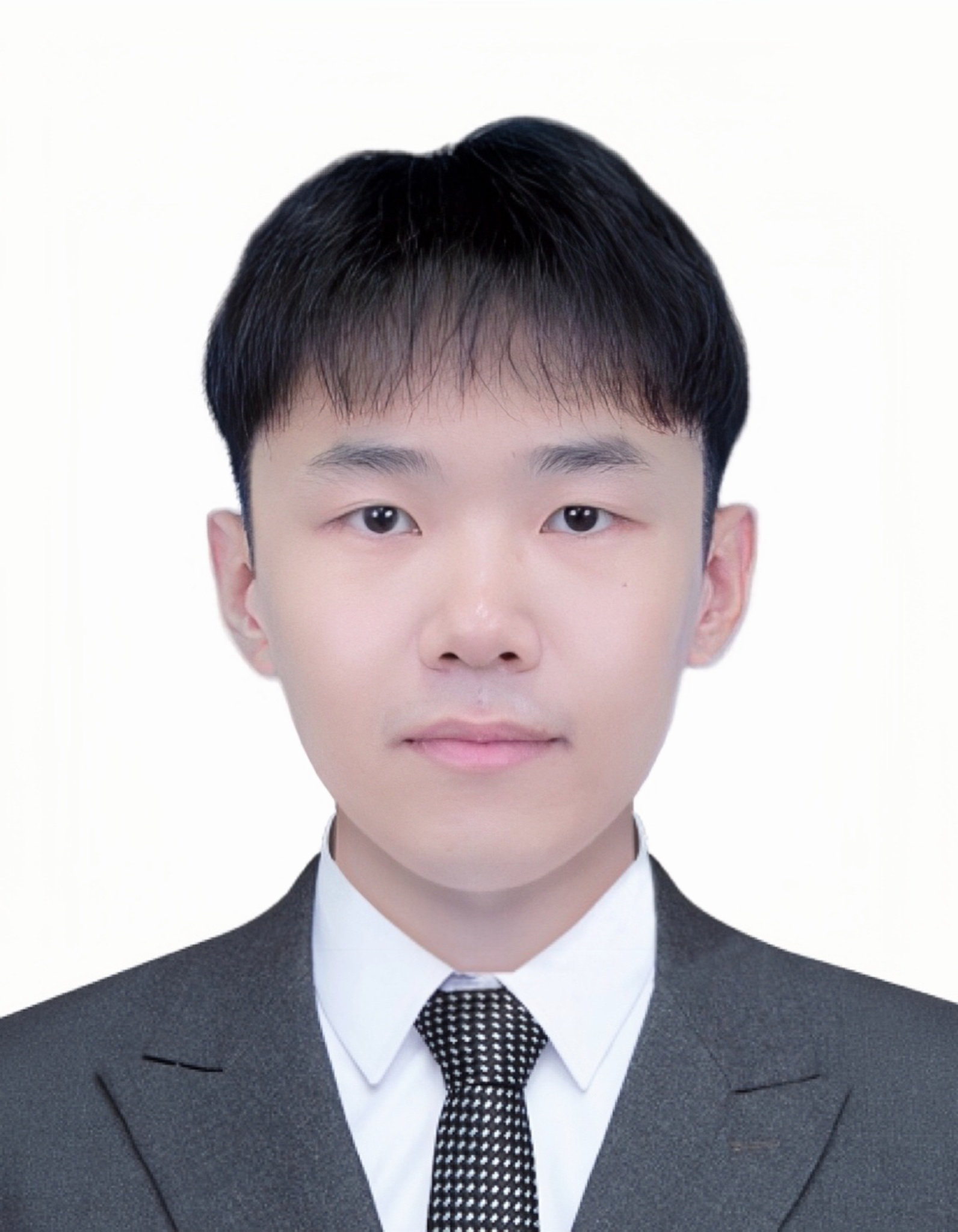}}]{Dongze Wu} received his B.Sc. degree in mathematics and applied mathematics from The Chinese University of Hong Kong, Shenzhen, China, in 2023. He is currently pursuing the M.Sc. degree in Statistical Science at the University of Oxford. His research interests include wireless communications, statistical signal processing, and neuroscience.
\end{IEEEbiography}

\begin{IEEEbiography}[{\includegraphics[width=1in,height=1.25in,clip,keepaspectratio]{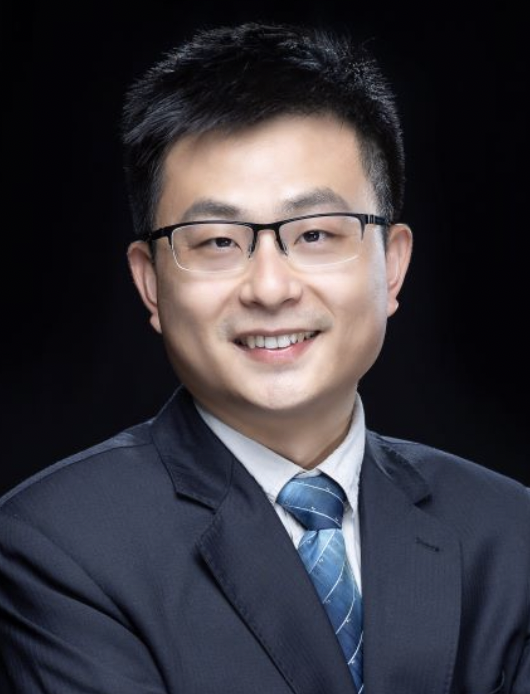}}]{Feng Yin} received his B.Sc. degree from Shanghai Jiao Tong University, China, and his M.Sc. and Ph.D. degrees from Technische Universitaet Darmstadt, Germany. From 2014 to 2016, he was a Marie-Curie Postdoc researcher with Ericsson Research, Linkoping, Sweden. Since 2016, he has been with The Chinese University of Hong Kong, Shenzhen as assistant professor. His research interests include statistical signal processing, Bayesian learning and optimization, and sensory data fusion. He has published around 90 top-tier journal papers and flag-ship conference papers, more than 20 US and China patents/communication standards. He was a recipient of the Chinese Government Award for Outstanding Self-Financed Students Abroad in 2013 and the Marie Curie Young Scholarship from the European Union in 2014. He was the finalist for the IEEE CAMSAP conference best paper award in 2013 and received the best paper award of Springer ICSINC conference in 2022. He is an IEEE senior member and currently serves as the Associate Editor for the Elsevier Signal Processing Journal.
\end{IEEEbiography}
\begin{IEEEbiography}[{\includegraphics[width=1in,height=1.25in,clip,keepaspectratio]{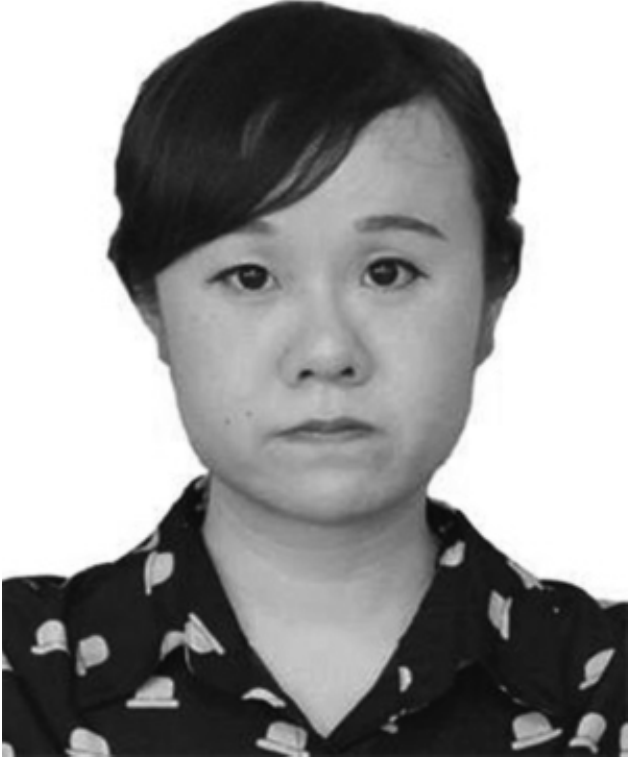}}]{Qinglei Kong} received the B.Eng. degree in communication engineering from the Harbin Institute of Technology, Harbin, China, in 2012, the M.Eng. degree in electronic and information engineering from the Shenzhen Graduate School, Harbin Institute of Technology, Shenzhen, China, in 2015, and the Ph.D. degree from the School of Electrical and Electronics Engineering, Nanyang Technological University, Singapore, in 2018. She is currently working as an Assistant Professor with the Institute of Space Science and Applied Technology, Harbin Institute of Technology (Shenzhen). She used to work with Cyber Security Cluster, Institute for Infocomm Research, Singapore, and Tencent Security, Shenzhen as a Research Scientist. She was a Postdoctoral Researcher with the Chinese University of Hong Kong, Shenzhen. Her research interests include applied cryptography, blockchain, VANET, and game theory.
\end{IEEEbiography}

\begin{IEEEbiography}[{\includegraphics[width=1in,height=1.25in,clip,keepaspectratio]{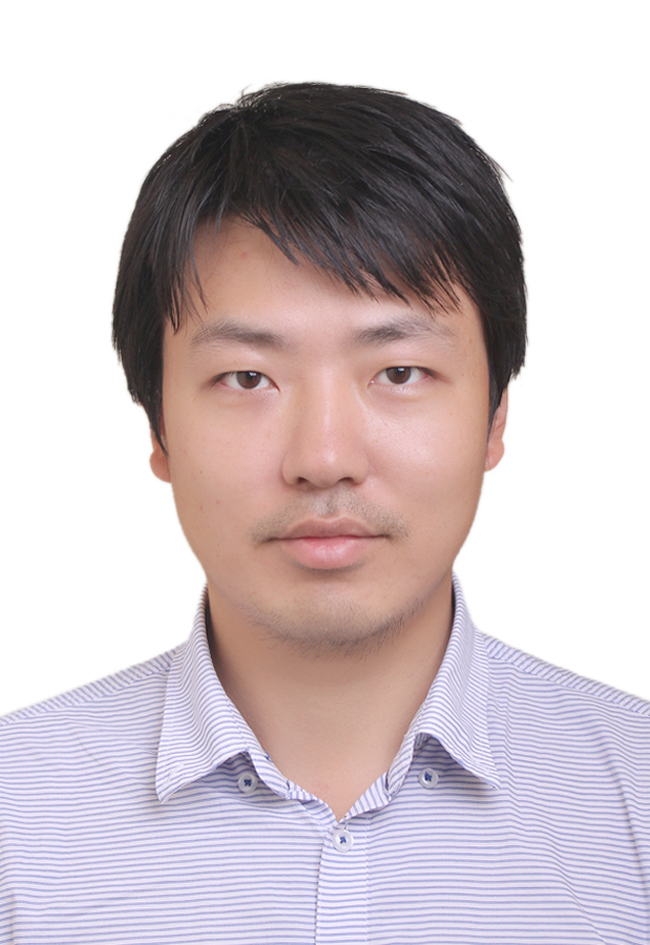}}]{Lexi Xu} received M.S. and Ph.D. degrees from Beijing University of Posts and Telecommunications, Beijing, China, and Queen Mary University of London, London, United Kingdom, in 2009 and 2013, respectively. From 2013 to 2020, he was a senior engineer at Network Technology Research Institute, China United Network Communications Corporation (China Unicom). Since August 2020, Dr. Xu is now a senior engineer at Research Institute, China Unicom. He is also a China Unicom delegate in ITU, ETSI, 3GPP, CCSA. He also serves as a professor (part-time) at Beijing University of Posts and Telecommunications. His research interests include big data, self-organizing networks, satellite system, radio resource management in wireless system, etc.
\end{IEEEbiography}
\begin{IEEEbiography}[{\includegraphics[width=1in,height=1.25in,clip,keepaspectratio]{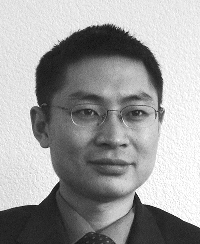}}]{Shuguang Cui} (S’99-M’05-SM’12-F’14) received his Ph.D in Electrical Engineering from Stanford University, California, USA, in 2005. Afterwards, he has been working as assistant, associate, full, Chair Professor in Electrical and Computer Engineering at the Univ. of Arizona, Texas A\&M University, UC Davis, and CUHK at Shenzhen respectively. He has also served as the Executive Dean for the School of Science and Engineering at CUHK, Shenzhen, the Executive Vice Director at Shenzhen Research Institute of Big Data, and the Director for Future Network of Intelligence Institute (FNii). His current research interests focus on the merging between AI and communication neworks. He was selected as the \textbf{Thomson Reuters Highly Cited Researcher} and listed in the \textbf{Worlds’ Most Influential Scientific Minds} by ScienceWatch in 2014. He was the recipient of the \textbf{IEEE Signal Processing Society 2012 Best Paper Award}. He has served as the general co-chair and TPC co-chairs for many IEEE conferences. He has also been serving as the area editor for IEEE Signal Processing Magazine, and associate editors for IEEE Transactions on Big Data, IEEE Transactions on Signal Processing, IEEE JSAC Series on Green Communications and Networking, and IEEE Transactions on Wireless Communications. He has been the elected member for IEEE Signal Processing Society SPCOM Technical Committee (2009$\sim$2014) and the elected Chair for IEEE ComSoc Wireless Technical Committee (2017$\sim$2018). He is a member of the Steering Committee for IEEE Transactions on Big Data and the Chair of the Steering Committee for IEEE Transactions on Cognitive Communications and Networking. He is also the Vice Chair of the IEEE VT Fellow Evaluation Committee and a member of the IEEE ComSoc Award Committee. \textbf{He was elected as an IEEE Fellow in 2013, an IEEE ComSoc Distinguished Lecturer in 2014, and IEEE VT Society Distinguished Lecturer in 2019}. In 2020, he won the IEEE ICC best paper award, ICIP best paper finalist, the IEEE Globecom best paper award. In 2021, he won the IEEE WCNC best paper award. In 2023, he won the IEEE Marconi Best Paper Award, got elected as a Fellow of  Canadian Academy of Engineering, and starts to serve as the Editor-in-Chief for IEEE Transactions on Mobile Computing. 
\end{IEEEbiography}
\end{document}